\documentclass[english,aps,pre,twocolumn,superscriptaddress,showpacs]{revtex4}

\usepackage[utf8]{inputenc}	

\usepackage{graphicx}
\usepackage{color}
\usepackage{subfigure}
\usepackage{amsmath}
\usepackage{amssymb}

\newcommand{\eq}[1]{Eq.~(#1)}
\newcommand{\fig}[1]{Fig.~#1}
\newcommand{\mean}[1]{\ensuremath{\langle{#1}\rangle}}
\newcommand{\abs}[1]{\ensuremath{\vert{#1}\vert}}

\graphicspath{{figures/}}

\begin{document}

\title{Chaos Pass Filter: Linear Response of Synchronized Chaotic Systems}

\author{Steffen Zeeb}
\email{steffen.zeeb@physik.uni-wuerzburg.de}
\affiliation{Institute of Theoretical Physics, University of W\"urzburg, Am Hubland, 97074 W\"urzburg, Germany}
\author{Johannes Kestler}
\affiliation{Institute of Theoretical Physics, University of W\"urzburg, Am Hubland, 97074 W\"urzburg, Germany}
\author{Ido Kanter}
\affiliation{Department of Physics, Bar-Ilan University, Ramat-Gan 52900, Israel}
\author{Wolfgang Kinzel}
\affiliation{Institute of Theoretical Physics, University of W\"urzburg, Am Hubland, 97074 W\"urzburg, Germany}

\date{\today}

\begin{abstract}

The linear response of synchronized time-delayed chaotic systems to small external perturbations, i.e., the phenomenon of chaos pass filter, is investigated for iterated maps. 
The distribution of distances, i.e., the deviations between two synchronized chaotic units due to external perturbations on the transfered signal, is used as a measure of the linear response. It is calculated numerically and, for some special cases, analytically. Depending on the model parameters this distribution has power law tails in the region of synchronization leading to diverging moments of distances. This is a consequence of multiplicative and additive noise in the corresponding linear equations due to chaos and external perturbations.
The linear response can also be quantified by the bit error rate of a transmitted binary message which perturbs the synchronized system. The bit error rate is given by an integral over the distribution of distances and is calculated analytically and numerically. It displays a complex nonmonotonic behavior in the region of synchronization. For special cases the distribution of distances has a fractal structure leading to a devil's staircase for the bit error rate as a function of coupling strength. The response to small harmonic perturbations shows resonances related to coupling and feedback delay times. A bi-directionally coupled chain of three units can completely filtered out the perturbation. Thus the second moment and the bit error rate become zero. 

\end{abstract}

\maketitle

\section{Introduction}
\label{sec:introduction}

Chaotic systems which are coupled to each other, can synchronize to a common chaotic trajectory \cite{Pecora:1990,Pikovsky:Buch,Schuster:2005,Boccaletti:2002}. 
Even for large delay times of the exchanged signals the dynamic systems can completely synchronize without any time shift \cite{Klein:2006:73,Fischer:2006:PRL,Sivaprakasam:2003,Lee:2006:JOSAB,Kinzel-2009,Englert-2010}.
This phenomenon is presently attracting a lot of attention, partly because of its counter-intuitive fundamental aspect of nonlinear dynamics, 
and partly because of its potential for secure communication with chaotic signals \cite{Cuomo:1993,Kocarev:1995,VanWiggeren:1998,Cuomo:1993:IEEE}. 
In fact, broadband communication with synchronized chaotic semiconductor lasers has recently been demonstrated over 120 km in a public fiber network \cite{Argyris:2005}. 

Several methods for secure chaos communication over a public channel have been suggested \cite{KinzelHandbook}. One of these utilizes a phenomenon which was coined ``chaos pass filter'' \cite{Murakami:2005,Fischer:2000}.
A chaotic receiver which is driven by the chaotic trajectory of a sender plus a (small) message responds essentially to the trajectory but not to the message. Thus the chaotic system filters out any perturbation and it is possible for the receiver to recover the message by subtracting its own dynamics from the incoming chaotic signal. 

Chaos pass filter is observed in experiments on electronic circuits and lasers and in simulations of chaotic systems. But the underlying physical process is still not well understood. When coupled chaotic units synchronize, their chaotic trajectories are attracted towards the synchronization manifold, thus the synchronized system's dynamics is restricted to the synchronization manifold. For stable chaos synchronization any random perturbation perpendicular to the manifold will exponentially decay to zero. However, this does not necessarily mean that a permanent perturbation like a message is damped as well. 

In this paper the phenomenon of chaos pass filter, i.e.\ the linear response of synchronized chaotic systems to an external perturbation is investigated. Research on linear response of general chaotic systems has been done before, mainly in the context of linear stochastic systems with multiplicative and additive noise \cite{Cessac:2007,Nakao:1998}. 
Here, we use iterated maps in order to obtain analytical results.
In some respect, for example for phase synchronization \cite{Rosenblum:1996}, chaotic maps have different properties than chaotic flows. But with respect to complete synchronization, maps and flows
are very similar and many of the obtained results are also observed in numerical simulations of chaotic differential equations.

This article is structured as follows: In section \ref{sec:theModels}, the model of two coupled chaotic units is introduced and the linearized system's equations are derived. Section \ref{sec:linearResponse} defines quantities which measure the linear response to noisy signals, such as the second moment and the bit error rate. The previously introduced model is studied extensively. Large excursions off the synchronization manifold are observed which lead to a continuum of diverging moments. Bit error rates, devil's stair cases, resonances and Lyapunov spectra are calculated. In section \ref{sec:ComplicatedModels} the investigation is extended to more complicated models, such as a chain of three units and a four units network. Finally, the last section summarizes and discusses our results.

\section{The Two Units Setup}
\label{sec:theModels}

\begin{figure}
\centering
 	\includegraphics[width=0.8\columnwidth]{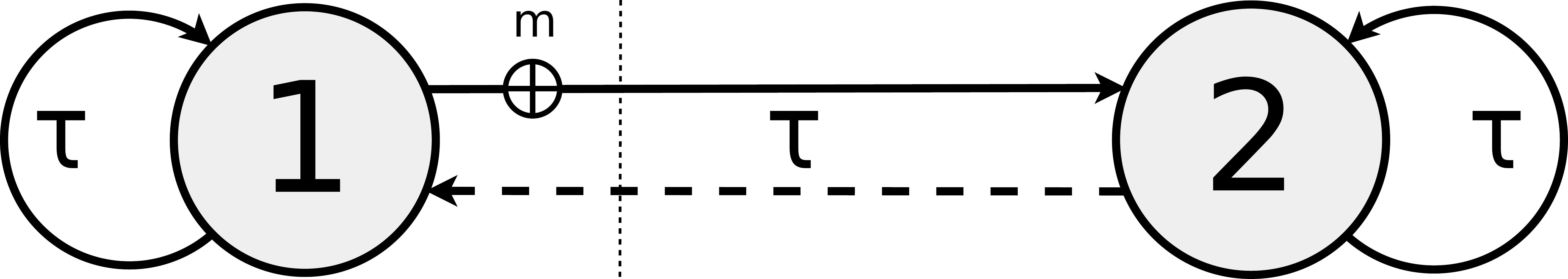}
	\caption{Setup of two coupled chaotic units with either uni- or bi-directional (dashed line) coupling. A perturbation $m$ is added to the exchanged signal at unit 1, i.e., the sender. The transmitted signal has a time delay $\tau$.}
	\label{fig:Setup}
\end{figure}

The simplest model for a coupled interacting chaotic system comprises of just two chaotic units which can either be uni- or bi-directionally coupled by a function of their internal variables. The exchanged signal usually has some time delay $\tau$ and in order for the bi-directional setup to be able to synchronize we in general include some self-feedback having the same time delay. 
The external perturbation $m$ is added to the transmitted signal at the sender as depicted in \fig{\ref{fig:Setup}}. The unidirectional configuration has been realized in a communication network over 120 km \cite{Argyris:2005}. 
In terms of chaos communication, where such a setup is called chaos masking \cite{KinzelHandbook}, one can think of the perturbation being a message which shall be secretly transmitted on top of a chaotic carrier signal. The message is small compared to the carrier signal and for our purpose its content may be considered as random noise. Hence we are, in the following, interested in the linear response of the receiver to noise.

Note that for chaos communication a uni-directional setup is not secure since an eavesdropper who, for example, knows all the details can use an identical copy of the receiver, synchronize it as well and extract the message. In contrast, two chaotic units which interact by bi-directional transmission have an advantage over an attacker driven by a uni-directional signal \cite{Kanter-2008}.

The dynamics of these two chaotic systems is given by the following sets of iterated equations
\begin{itemize}
	\item uni-directional setup
		\begin{align} 
			x_{t+1} &= (1 - \epsilon) f(x_t) + \epsilon f(x_{t - \tau}) \notag \\
			y_{t+1} &= (1 - \epsilon) f(y_t)  \notag \\
				& \quad + \epsilon \kappa f(y_{t - \tau}) + \epsilon (1 - \kappa) f(x_{t - \tau} + m_{t - \tau}) \: ,
		\label{eq:setupA}
		\end{align} 
	\item bi-directional setup
		\begin{align} 
			x_{t+1} &= (1 - \epsilon) f(x_{t})  \notag \\
				& \quad + \epsilon \kappa f(x_{t - \tau}) + \epsilon (1 - \kappa) f(y_{t - \tau} ) \notag \\
			y_{t+1} &= (1 - \epsilon) f(y_{t}) \notag \\
				& \quad + \epsilon \kappa f(y_{t - \tau}) + \epsilon (1 - \kappa) f(x_{t - \tau} + m_{t - \tau}) \: ,
		\label{eq:setupB}
		\end{align}
\end{itemize}
where $t$ is a discrete time step. The parameter $\epsilon$ measures the contribution of all delay terms while parameter $\kappa$ measures the relative strength of the self-feedback. They are chosen such that the dynamics,  $x_t$ and $y_t$, always stays in the unit interval $[0,1]$. 
For the function $f(x)$ one of the following discrete maps is used:
\begin{itemize}
	\item Bernoulli map 
	\begin{align}
		f(x) = (a \, x) \mod 1
	\end{align} 
	\item Tent map
	\begin{align}
		f(x) =   \left\{ 
		\begin{array}{ll} 
			\frac{1}{a} \, x & \mathrm{for} \: 0 \leq x < a  \\
			\frac{1}{1-a} \, (1-x) & \mathrm{for} \: a \leq x \leq 1
		\end{array} \right. 
	\end{align}
	\item Logistic map
		\begin{align} \label{eq:logistischeAbbildung}
 			f(x) = a\,x (1 - x) 
		\end{align} 
\end{itemize}

\begin{figure} 
\centering
	\includegraphics[width=0.7\columnwidth]{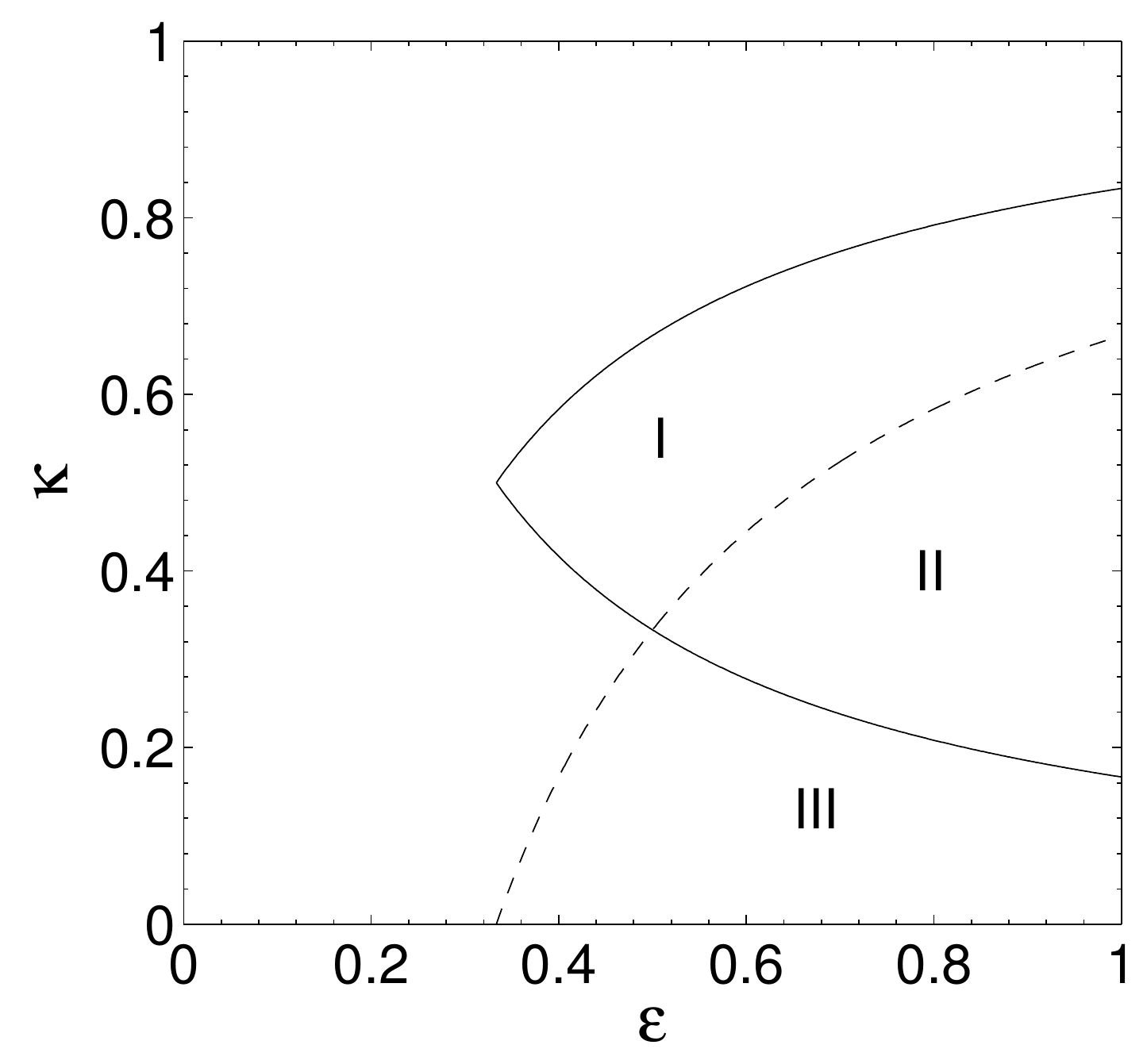}
	\caption{Phase diagram for two coupled Bernoulli units with $a=1.5$. System is completely synchronized in regions II + III (- -) for the the uni-directional setup and in regions  I + II (-) for the bi-directional setup,  see \eq{\ref{eq:gebietA}} and \eq{\ref{eq:gebietB}}.}
	\label{fig:SyncArea}
\end{figure}

Without noise, $m_t = 0$, the spectra of Lyapunov exponents and hence the phase diagram of synchronization can be calculated analytically for the Bernoulli system in the limit of large delay, $\tau \to \infty$ \cite{Lepri1994,Englert-2011}. 
The parameter range for which the system in the stationary state synchronizes completely, i.e., $x_t = y_t$, is depicted in \fig{\ref{fig:SyncArea}} and is given by the inequality 
\begin{itemize}
	\item uni-directional setup
	\begin{align}
		\kappa < \frac{1-a(1-\epsilon)}{a \epsilon} \: ,
		\label{eq:gebietA}
	\end{align}
	\item bi-directional setup
	\begin{align}
		\frac{a-1}{2 a \epsilon} < \; & \kappa < \frac{1+ 2 a \epsilon - a}{2 a \epsilon}  \: .
		\label{eq:gebietB}
	\end{align}
\end{itemize}

The effect of small noise can be calculated by linearizing equations (\ref{eq:setupA}) and (\ref{eq:setupB}) in the vicinity of the synchronization manifold $x_t = y_t =s_t$. The noise leads to a small deviation $d_t = y_t - x_t$ which is determined by the linear equations
\begin{itemize}
	\item uni-directional setup
		\begin{align}
			d_{t+1} &= (1-\epsilon) f'_t d_t + \epsilon \kappa f'_{t - \tau} d_{t - \tau}  \notag \\
				& \qquad + \epsilon (1 - \kappa) f'_{t - \tau} m_{t - \tau} \: ,
		\label{eq:linA}
		\end{align}
	\item bi-directional setup
		\begin{align} 
			d_{t+1} & = (1 - \epsilon) f'_t d_t + \epsilon \kappa f'_{t - \tau} d_{t - \tau}  \notag \\
				& \qquad  + \epsilon (1 - \kappa) f'_{t - \tau} \, (m_{t - \tau}  - d_{t - \tau}) \notag \\
			 & = (1 - \epsilon) f'_t d_t + (2 \kappa - 1) \epsilon f'_{t - \tau} d_{t - \tau} \notag \\
				& \qquad  + \epsilon (1 - \kappa) f'_{t - \tau} m_{t - \tau}  
		\label{eq:linB}
		\end{align}
\end{itemize}
where $f'_{t}$ is the derivative of $f(x)$ at the synchronized trajectory $s_t$. 
These equations determine the region of synchronization. Without noise $d_t$ decays to zero for stable synchronization and increases exponentially otherwise. Note that in this case, i.e., without any noise present, these equations can be solved analytically for Bernoulli units, where $f'_t = a = const$, and determine the phase diagram of \fig{\ref{fig:SyncArea}}.

\section{Linear Response}
\label{sec:linearResponse}

We want to understand how a synchronized chaotic system responds to small perturbations, i.e., the linear response of the system. In the following we analyze the linear response by means of the moments of the distribution of $d_t$, the bit error rate of a transmitted random message and resonances of harmonic perturbations.

\subsection{Moments}

Without an external perturbation the coupled system can synchronize completely, see \fig{\ref{fig:SyncArea}}. With noise however the deviation from the synchronization manifold $d_t$ does not decay to zero but has a distribution around zero.
We are interested in the stationary distribution which develops after a transient time and which is caused by the transverse Lyapunov exponents. This distribution is characterized by its moments which are defined as
\begin{align} 
	\chi_n = \lim_{\mean{\abs{m}} \to 0}  \frac{\mean{\abs{d}^n}}{\mean{\abs{m}^n}} \: ,
	\label{eq:chi}
\end{align} 
where $\mean{\ldots}$ means an average over the distribution of $m$ and $d$, respectively and $n$ denotes the order of the moment.

For investigating the moments we model the external perturbation $m_t$ as random numbers with a uniform distribution in the interval $[-M, M]$.

In order to obtain analytical results let us consider the simplest case, the uni-directional setup without any time delay, $\tau=0$. 
Using the substitution $\alpha=\epsilon(1-\kappa)$ the system reduces to a simple master-slave setup without self-feedback and the linearized equation \eqref{eq:linA} takes the simple form 
\begin{equation}
	d_{t+1} = (1 - \alpha) f'_t \, d_t + \alpha f'_t \, m_t \: .
	\label{eq:dtEinfach}
\end{equation} 
For Bernoulli maps $f'_t$ is a constant and for the other chaotic maps we assume $f'_t$ to be an uncorrelated random number. In the latter case \eq{\ref{eq:dtEinfach}} is a discrete linear stochastic equation with multiplicative and additive noise of the form 
\begin{align}
	x_{t+1} = \gamma_t x_t + \eta_t \: .
\end{align}
In the theory of discrete linear stochastic equation it is well known that such an equation may lead to stationary distributions which have a power law of the form \cite{Nakao:1998,Sornette:1998,Kuramoto:1997}
\begin{align}
	\rho(x) \sim \frac{1}{x^{\mu}} \: ,
\end{align}
where $\mu$ is a solution of the following equation
\begin{align}
	 \mean{\gamma^{\mu-1}} =1 \, .
	\label{eq:expoGamma}
\end{align}

\begin{figure}
\centering
	\includegraphics[width=0.85\columnwidth]{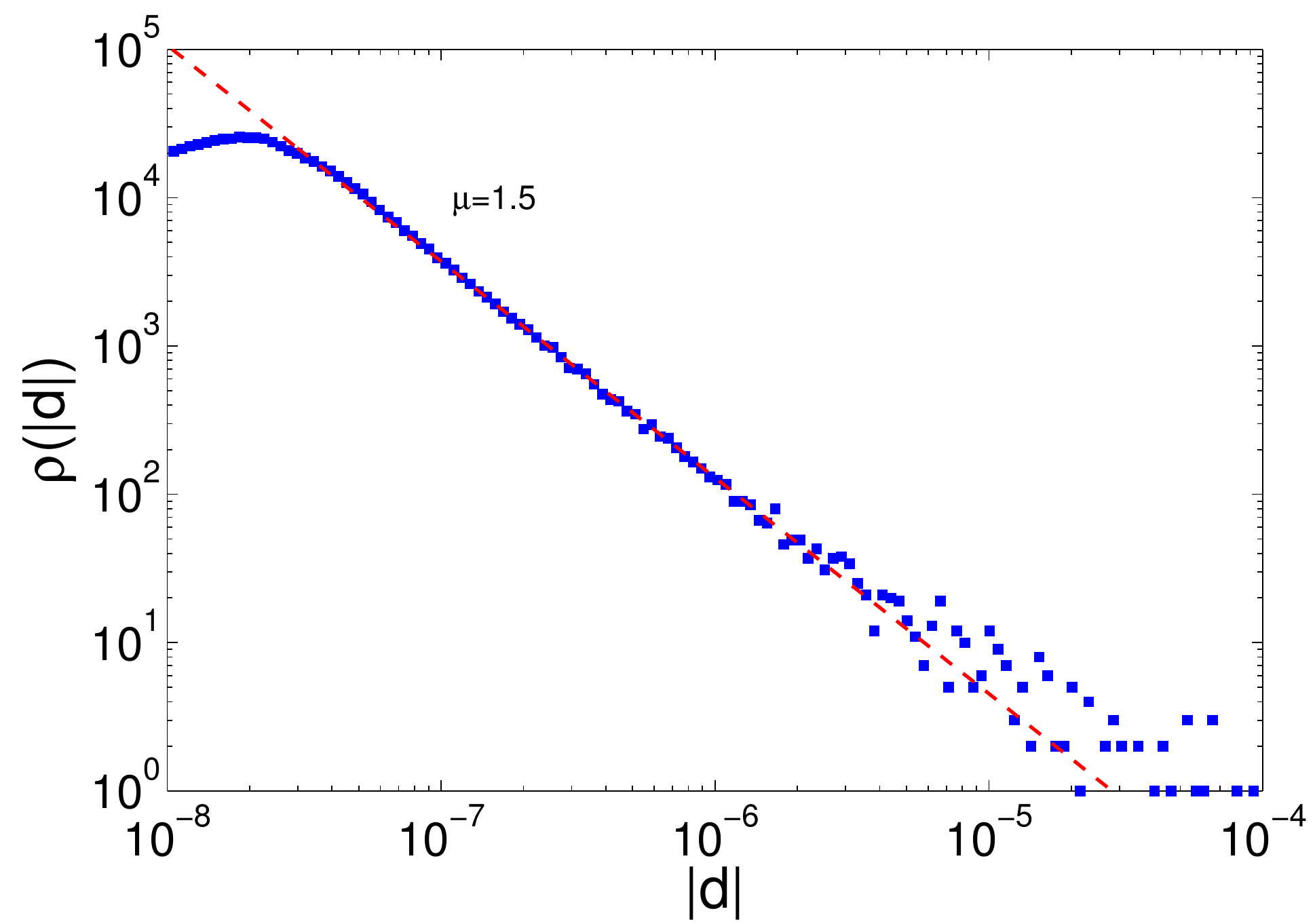}
	\caption{Distribution of $\abs{d}$ for two uni-directionally coupled logistic maps with $a=4$, $\kappa=0$, $\tau=0$ and $M=10^{-8}$ in a log-log plot. Dashed red line shows a power law fit. There is a cut-off from the power law behavior for small $d$ due to the additive noise which prevents the system from perfectly synchronizing.}
	\label{fig:loglog}
\end{figure}

As simulations show the distribution $\rho(d)$ may indeed follow a power law.
If the factor $(1 - \alpha) |f'_t|$ in \eq{\ref{eq:dtEinfach}} takes on values larger than one, $d_t$ can temporarily explode, i.e., the system can have very large excursions from the synchronized state, leading to a power law. For the logistic map with $a=4$, for which the distribution of $f'$ is given by 
\begin{equation}
	\rho(f') = \frac{1}{\pi \sqrt{16 - f'^2}} \: ,
	\label{eq:density:fs}
\end{equation}  
the maximum slope is $\abs{f'} = 4$ and hence we expect to find large excursions of $d_t$ for $\epsilon < \frac{3}{4}$.
\fig{\ref{fig:loglog}} shows a typical distribution of $\abs{d}$ for two uni-directionally coupled logistic maps. The distribution has a peak at values slightly larger than the maximum noise strength $M$, since the additive noise in the systems prevents it from being perfectly synchronized. For larger distances the distribution follows, to a good approximation, a power law.

We did not succeed in calculating the distribution $\rho(d)$ analytically, but the linear response can be derived.
Squaring \eq{\ref{eq:dtEinfach}} one finds for the second moment
\begin{equation} 
	\frac{\mean{d^2}}{\mean{m^2}} = \frac{\alpha^2 \mean{f'^2}}{1 - (1-\alpha)^2 \mean{f'^2}} \,.
	\label{eq:SecondMomentUni}
\end{equation} 
The second moment diverges when the denominator becomes zero meaning that the distribution $\rho(d)$ has developed a power law tail or in other words that rare but large excursions from the synchronization manifold occur.
For the Bernoulli map with $a=1.5$ the second moment diverges for $\alpha < \alpha_m = \frac{1}{3}$ which coincide with the synchronization transition given by
\begin{equation} \label{eq:syncBed}
	\ln (1 - \alpha) + \mean{ \ln |f'|} < 0 \: .
\end{equation}

\begin{figure}
\centering
	\includegraphics[width=0.9\columnwidth]{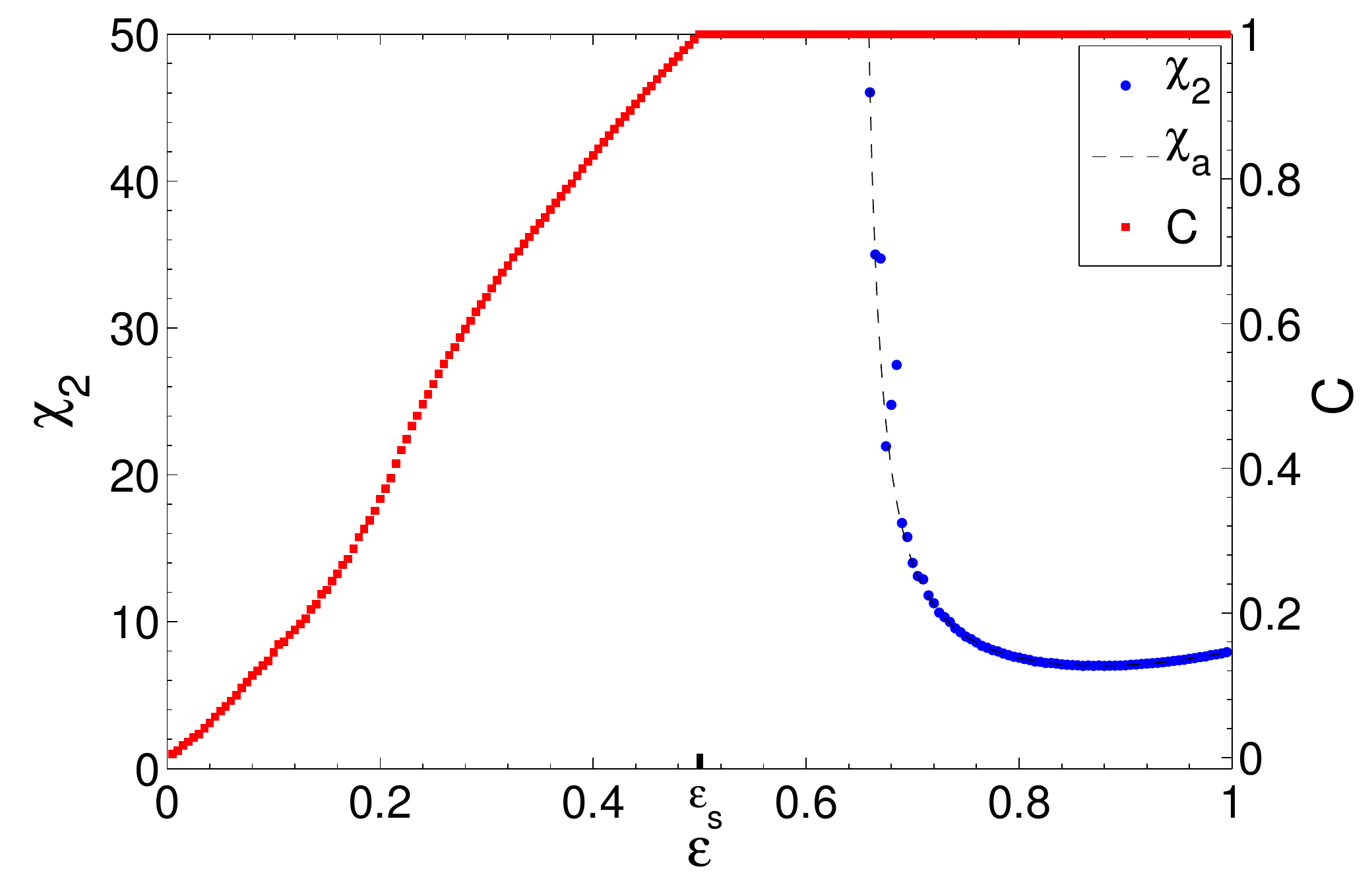}
\caption{Second moment $\chi_2$ (blue points) and cross correlation $C$ (red squares) for two uni-directionally coupled logistic maps  with $a=4$, $\tau=0$, $M=10^{-8}$ and $\kappa=0$ (such that $\alpha=\epsilon$) as a function of $\epsilon$. Dashed curve shows analytical results for $\chi_2$ whereas other results were obtained from simulations.}
\label{fig:SecondMomentUni}
\end{figure}

Without multiplicative noise all moments diverge at the same parameters.
For the logistic and the tent map, however, the second moment diverges already inside the region of synchronization. For the logistic map with $a=4$ and the mean values $\mean{{f'}^2} = 8$ and $\mean{\ln{\abs{f'}}}=\ln{2}$ which can be analytically computed from the distribution \eq{\ref{eq:density:fs}} we find the synchronization transition to be at $\alpha_s = \frac{1}{2}$ and the second moment to diverge at $\alpha_m = 1-\frac{1}{\sqrt{8}} \approx 0.646$. For the tent map with $a=0.4$ where $\mean{{f'}^2} = \frac{25}{6}$ and $\mean{\ln{\abs{f'}}} \approx 0.67$ we find $\alpha_s \approx 0.49$ and $\alpha_m = 1-\sqrt{\frac{6}{25}} \approx 0.51$.
\fig{\ref{fig:SecondMomentUni}} shows the second moment $\chi_2$ together with the cross-correlation $C$, which is a measure for the synchronization, for a system of coupled logistic maps. The numerically obtained results for the second moment agree with the analytical results \eq{\ref{eq:SecondMomentUni}} and one can see clearly that the second moment diverges even though the system is still synchronized.

\begin{figure}
\centering
	\includegraphics[width=0.85\columnwidth]{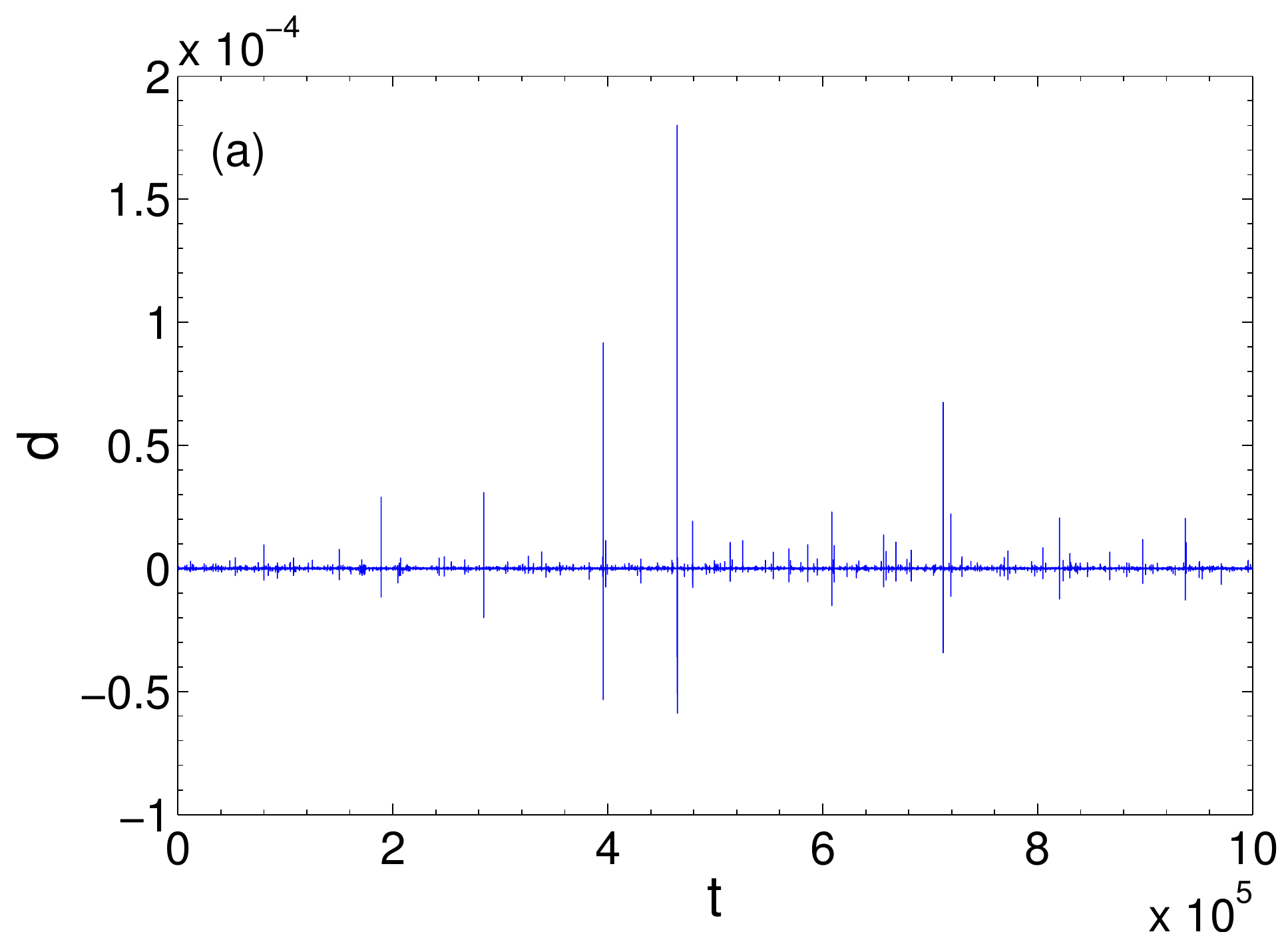}
	\includegraphics[width=0.85\columnwidth]{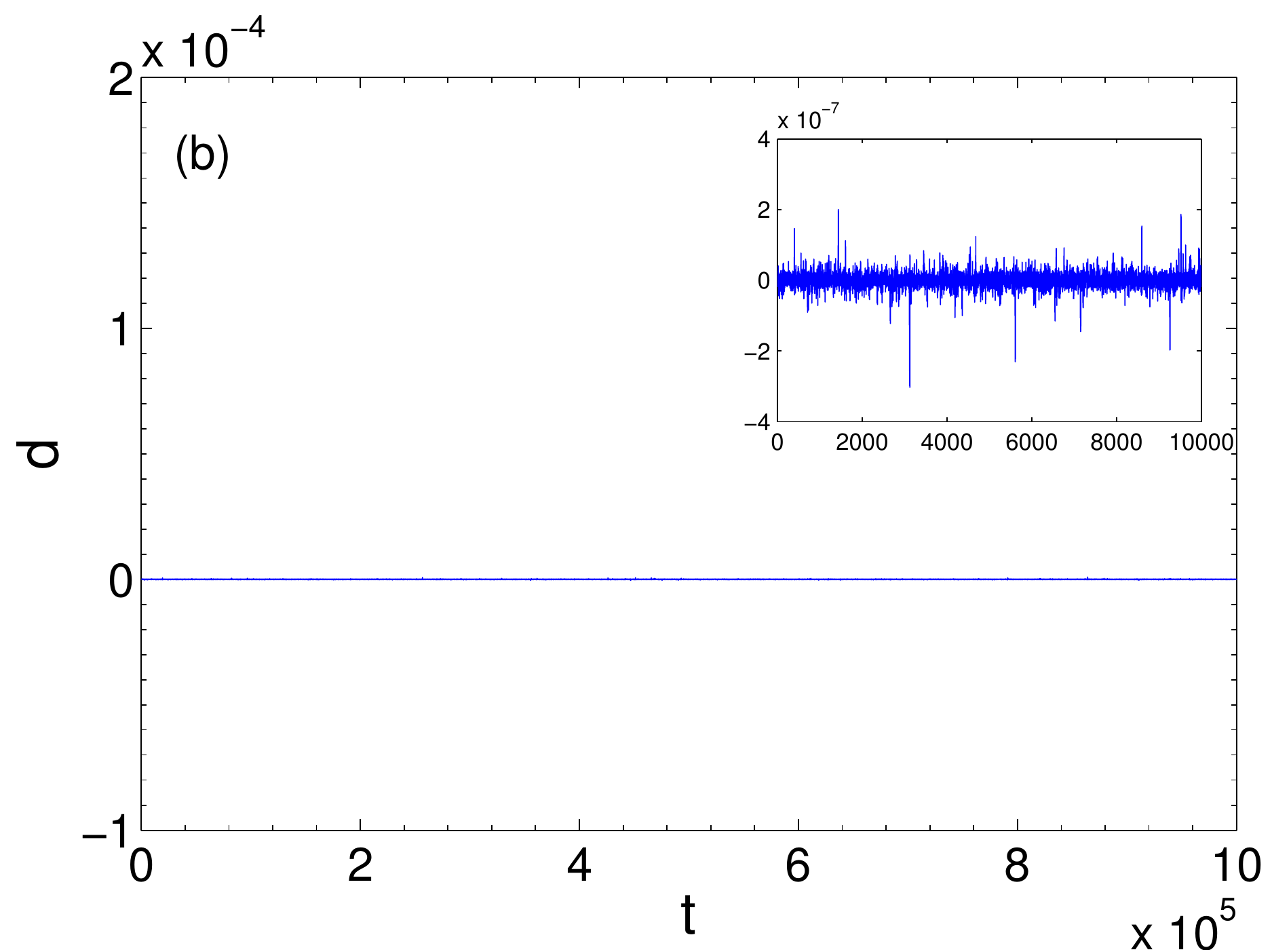}
	\caption{
Trajectory of $d_t$ for two uni-directionally coupled Logistic maps with $a=4$, $\kappa=0$, $\tau=0$, $M=10^{-8}$ and (a) $\epsilon = 0.6$ and (b) $\epsilon = 0.7$. In both cases the cross-correlation is $C=1$ but for (a) large excursions from the synchronization manifold occur and the second moment diverges, $\chi_2=4632$, whereas for (b) the second moment is finite, $\chi_2=14$. Inset in (b) shows that the deviations from the synchronization manifold is in the order of the magnitude of the noise.}	
	\label{fig:dt}
\end{figure}

\fig{\ref{fig:dt}} shows a trajectory of $d_t$ for logistic maps where the coupling parameters are once chosen such that the second moment exists and once that it diverges. In both cases the system is synchronized, i.e., $C=1$, but in case of a diverging second moment large excursions from the synchronization manifold occur. On the contrary, for a finite second moment the deviations from synchronization are mainly determined by the additive noise term in \eq{\ref{eq:dtEinfach}} and are of the magnitude of $M$.

Note that we obtain similar plots to \fig{\ref{fig:loglog}-\ref{fig:dt}} for coupled tent maps but due to the limited space we restrict the presentation of our results at this point to coupled logistic maps.

Power law tails of the distribution of a stochastic process with multiplicative and additive noise have been discussed in the context of chaos synchronization, before \cite{Pikovsky:Buch}. This phenomenon has been called ``on-off intermittency'' and was discussed in the vicinity of the synchronization transition. Here the phenomenon is triggered by the message transmitted via the chaotic signal, and it is observed deep inside the region of synchronization.

\begin{figure}
\centering
	\includegraphics[width=0.85\columnwidth]{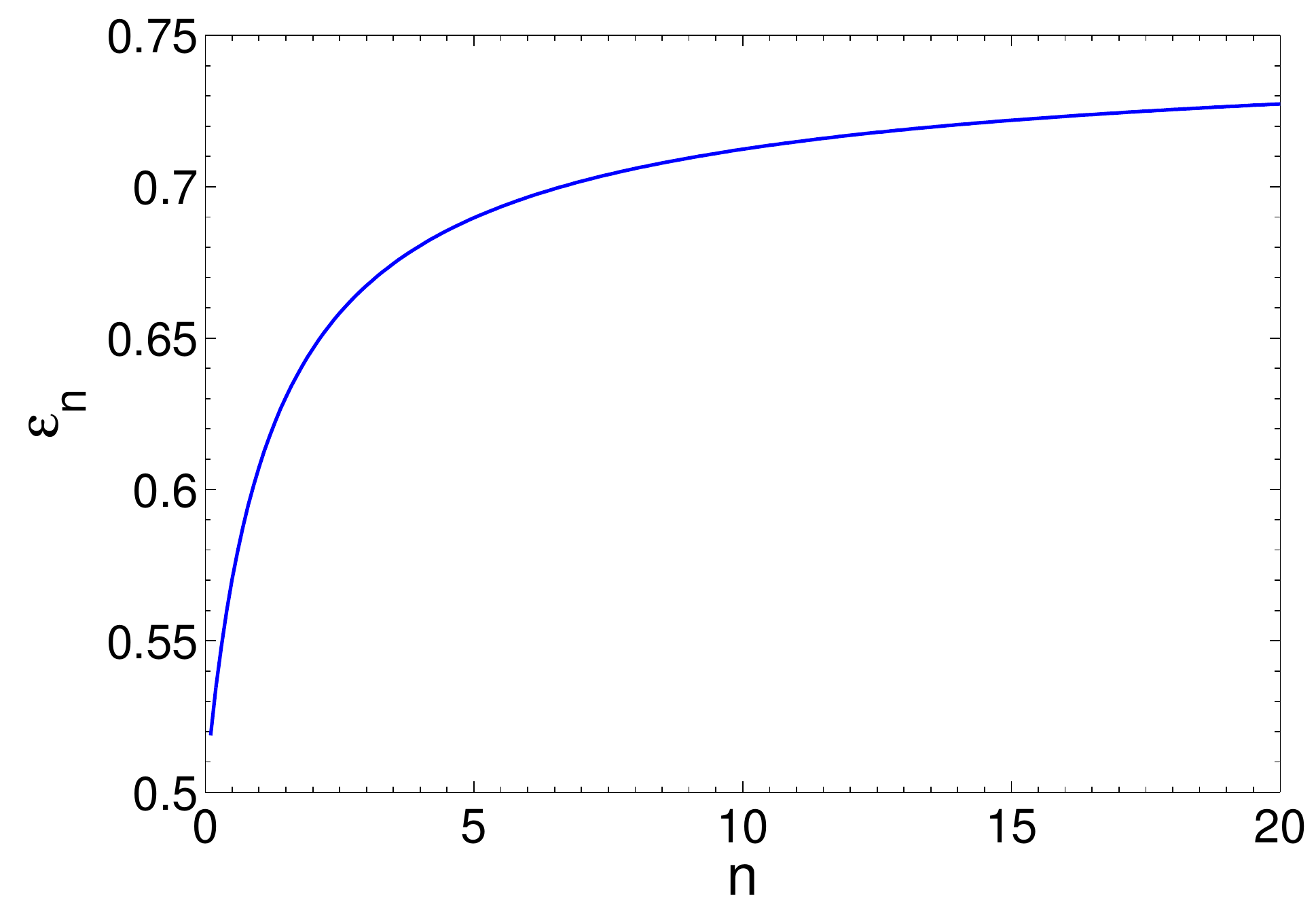}
	\caption{The threshold $\epsilon_n$ below which $\chi_n$ diverges for two uni-directionally coupled logistic maps with $a=4$, $\kappa=0$ and $\tau=0$, see \eq{\ref{eq:EpsNLog}}.}
	\label{fig:EpsNLog}
\end{figure}

The linear equation (\ref{eq:dtEinfach}) also determines higher moments of $p(d)$. By taking the $n$th power of \eq{\ref{eq:dtEinfach}} we find that $\chi_n$ diverges at a coupling $\epsilon_n$ given by
\begin{align}
	1 = (1 - \epsilon_n)^n \mean{f'^n} \: .
	\label{eq::EpsN}
\end{align} 
Hence, for Bernoulli maps all moments diverge at the synchronization threshold $\epsilon_s$. For logistic maps, however, all moments diverge at a different coupling strengths $\epsilon_n$ given by 
\begin{align}
	\epsilon_n = 1 - \frac{1}{4} \left(
	\frac{n \, \sqrt{\pi} \, \Gamma(\frac{n}{2})}{2 \, \Gamma(\frac{n+1}{2})}
	\right) ^ {\frac{1}{n}} \: .
	\label{eq:EpsNLog}
\end{align} 
Starting from a strongly synchronized state and decreasing $\epsilon$ the distribution $p(d)$ broadens and eventually follows a power law.
The different moments of the distribution successively start to diverge starting with the highest one $\chi_{\infty}$ and continuing with lower and lower ones until finally for $n \to 0$ the synchronization threshold is reached, where $\mean{\ln |d|}$ diverges. Figure \ref{fig:EpsNLog} shows $\epsilon_n$ as a function of $n$. Note that the broadening of the distribution by decreasing $\epsilon$ is related to a decreasing power law exponent $\mu$. Comparing equation \eqref{eq::EpsN} and \eqref{eq:expoGamma} we see that $\mu$ is related to the order of the moment $n$ by $\mu = n + 1$.

Up to now we have discussed the uni-directional setup only. The linearized equations for the bi-directionally coupled system without time delay are very similar and with the substitution $\alpha=\epsilon(1-\kappa)$ read
\begin{align} 
	d_{t+1} = (1 - 2 \alpha) f'_t d_t + \alpha f'_t m_t \: .
	\label{eq:dtZweifach}
\end{align} 
However this system is harder to analyze analytically since due to the mutual interaction the distribution of $f'$ is rendered. $\rho(f')$ does not correspond to the distribution of an isolated unit anymore as it was the case for the master-slave setup. Only for Bernoulli maps where $f'=const$ the distribution $\rho(f')$ is not affected by the mutual coupling.

\begin{figure}
\centering
	\includegraphics[width=0.9\columnwidth]{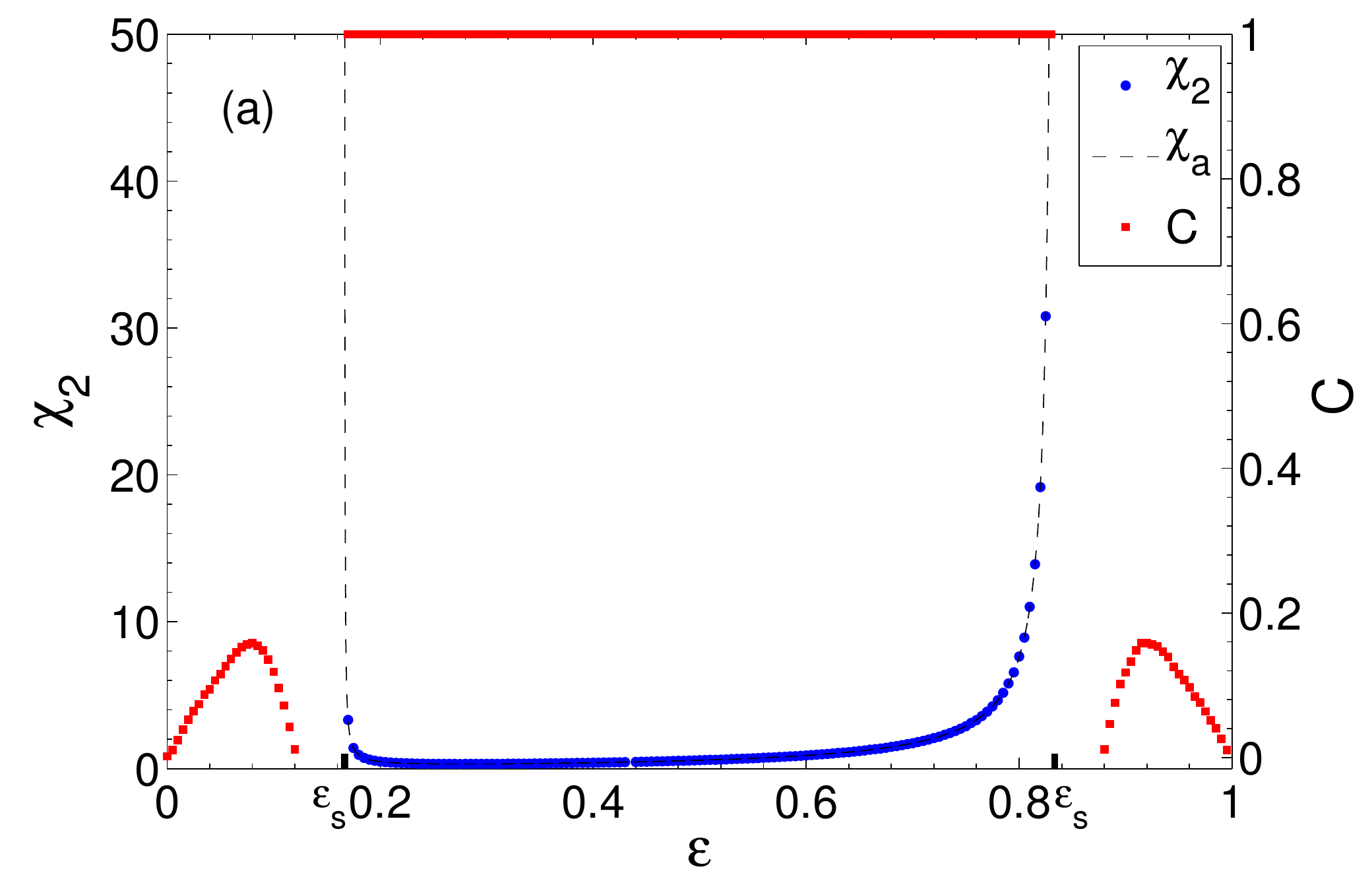}
	\includegraphics[width=0.9\columnwidth]{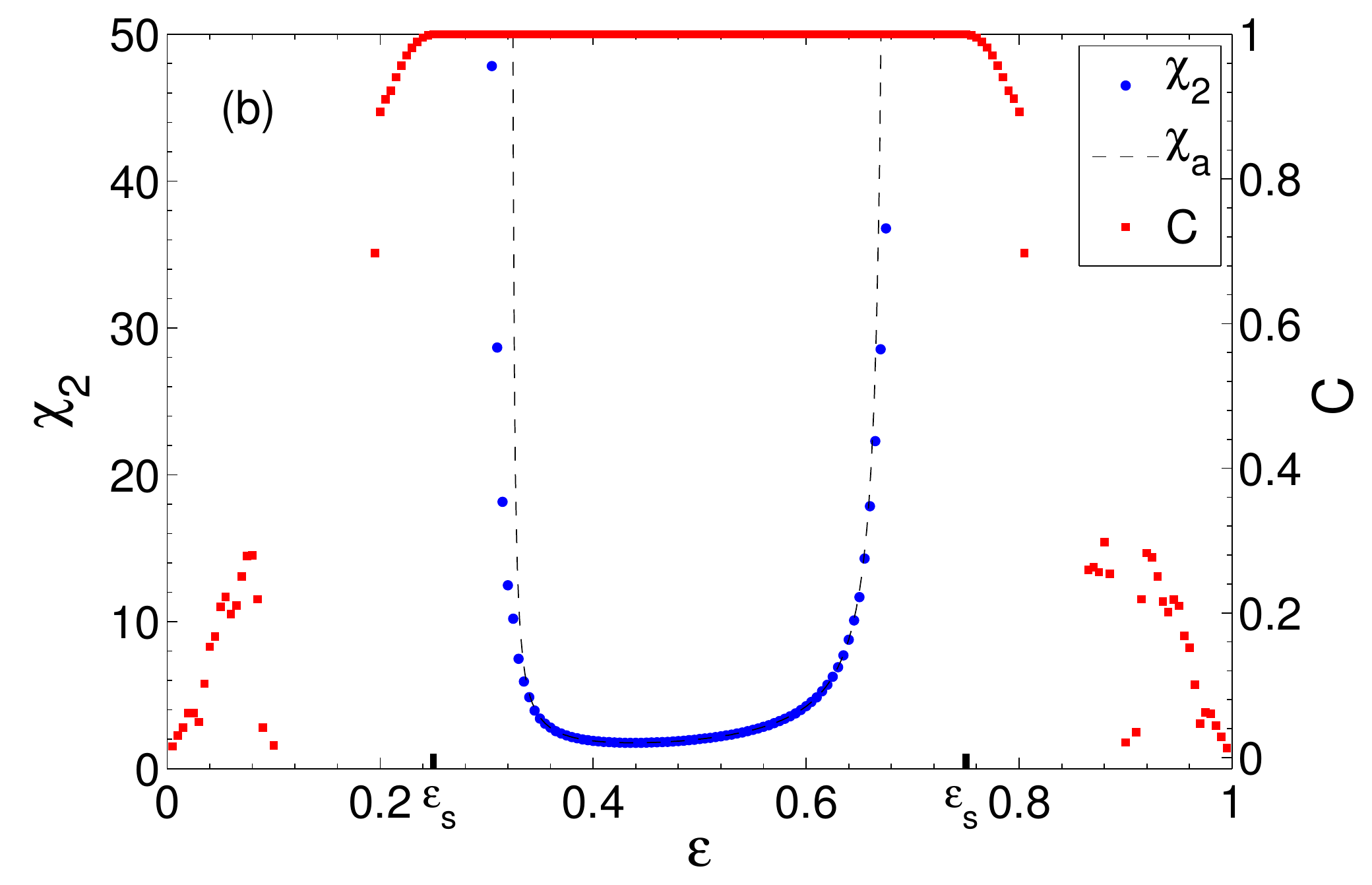}
\caption{Second moment $\chi_2$ (blue points) and cross correlation $C$ (red squares) for two bi-directionally coupled (a) Bernoulli and (b) logistic maps with $a=1.5$ and $a=4$, respectively, as a function of $\epsilon$. Other parameters are $\tau=0$, $M=10^{-8}$ and $\kappa=0$ (such that $\alpha=\epsilon$). Dashed curve shows analytical results for $\chi_2$ whereas other results were obtained from simulations.}
\label{fig:SecondMomentBi}
\end{figure}

\begin{figure}
\centering
	\includegraphics[width=0.49\columnwidth]{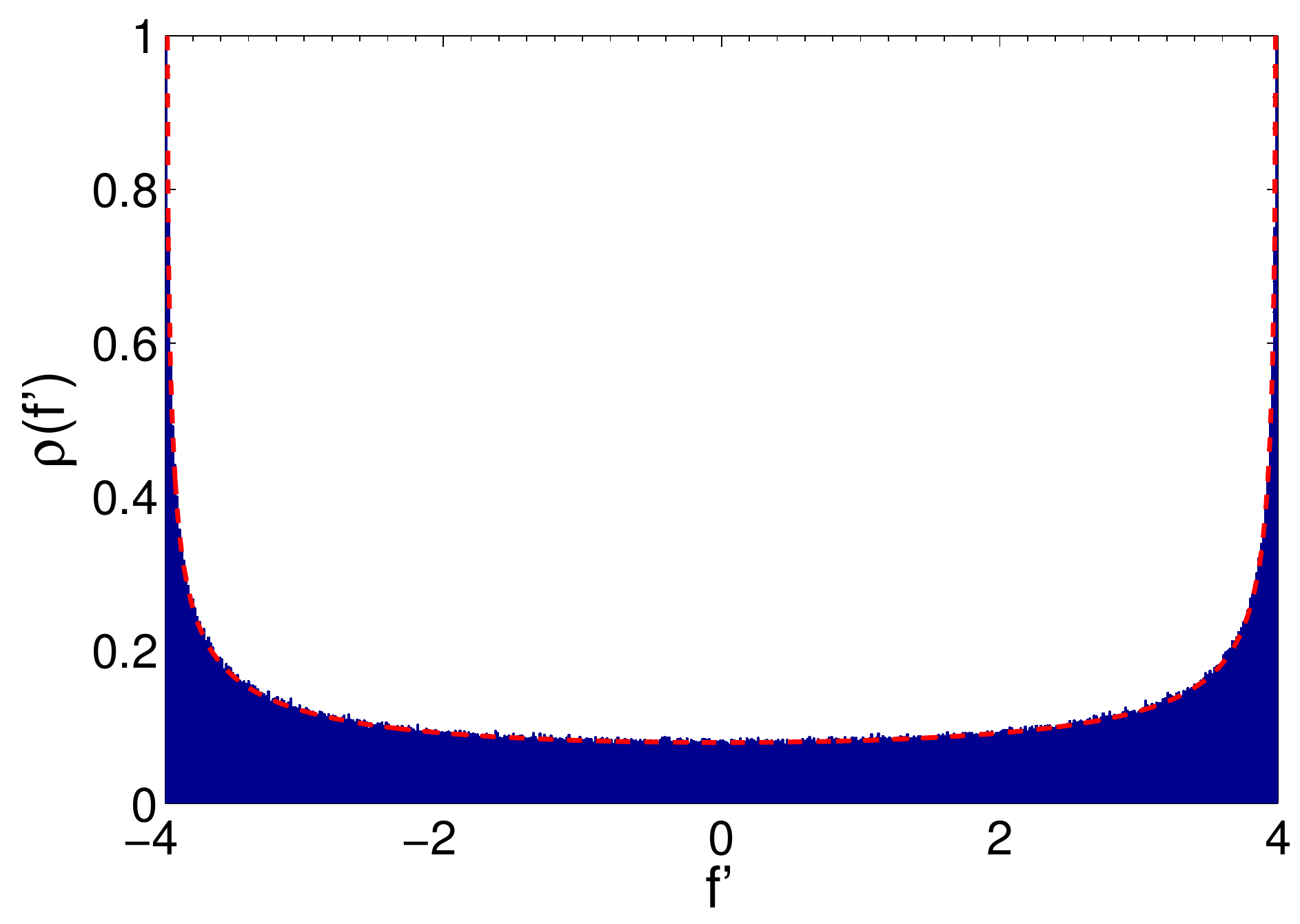}
	\includegraphics[width=0.49\columnwidth]{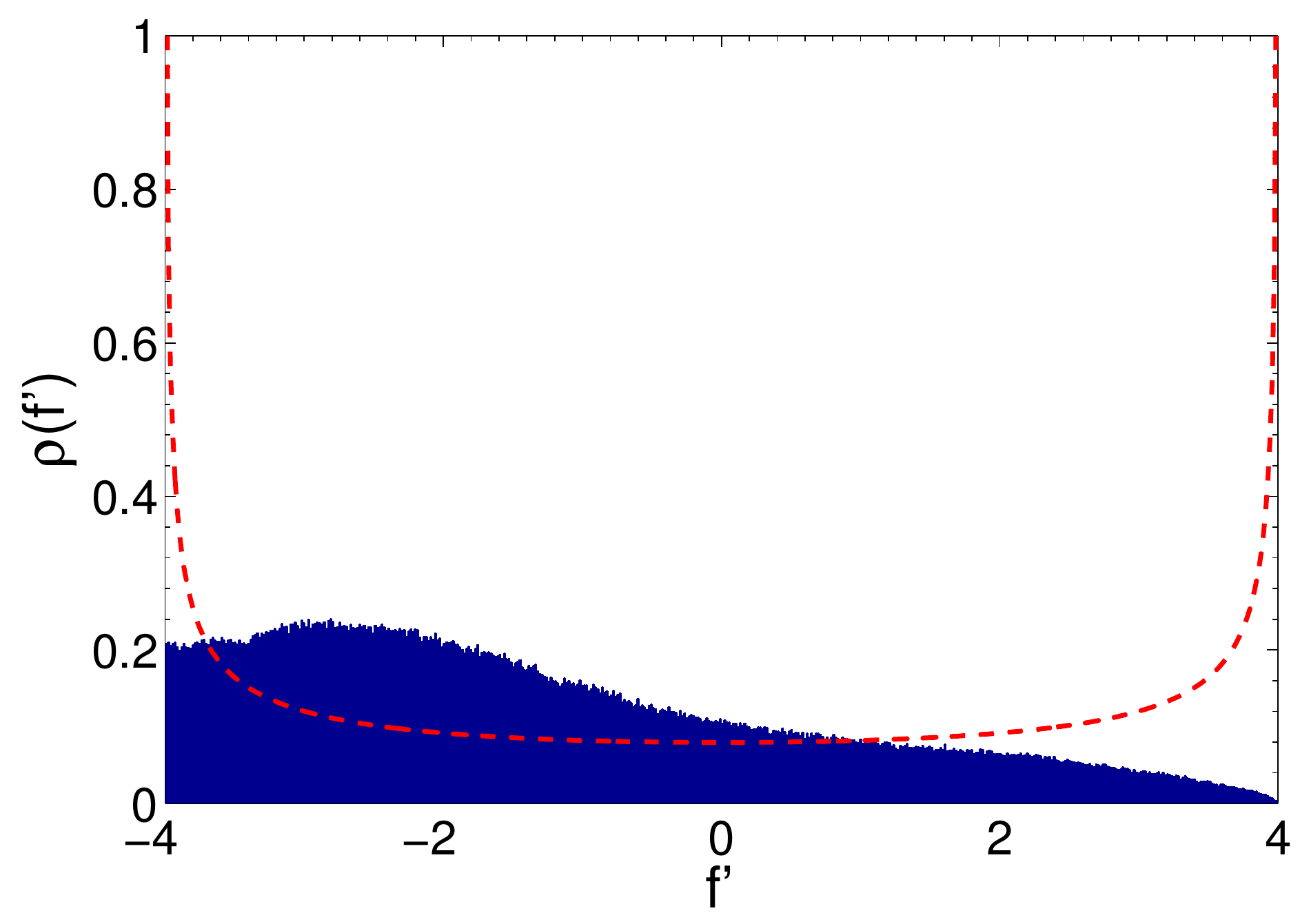}
\caption{Probability distribution $\rho(f')$ of the receiver of coupled synchronized logistic maps for (a) undelayed bi-directional setup with $\epsilon=0.4$ and $\tau=0$ and (b) time-delayed uni-directional setup with $\epsilon=0.7$ and $\tau=100$. Other parameters are $\kappa=0$ and $M=10^{-8}$. Dashed (red) line shows the distribution of a single logistic map given by equation \eqref{eq:density:fs}.}
\label{fig:DistFDash}
\end{figure}

The second moment is given by
\begin{align}
	\frac{\mean{d^2}}{\mean{m^2}} = \frac{\alpha^2 \mean{f'^2}}{1-(1-2\alpha)^2 \mean{f'^2}} \: ,
	\label{secondMoment_tau0_bi}
\end{align}
and it diverges if the denominator approaches zero. Although the mutual interaction renders the distribution of $f'$ and therefor we cannot compute the mean $\mean{f'^2}$ nor $\mean{\ln{\abs{f'}}}$ for the tent and logistic map, respectively, we obtain a surprisingly good agreement with numerical results if we use, as a first approximation, the distribution of a single unit for computing the means. Fig.~\ref{fig:SecondMomentBi} shows $\chi_2$ obtained from numerical simulations together with analytical results for Bernoulli and logistic maps. Only at the transition from finite to diverging second moments the simulations deviate slightly from the analytical results in case of logistic maps. This indicates that $\rho(f')$ is changed only moderately by the bi-directional coupling. Closer investigations of $\rho(f')$ have indeed confirmed that the distribution is hardly changed for $\tau=0$, see for example Fig.~\ref{fig:DistFDash} (a) which shows $\rho(f')$ for the receiver of two synchronized bi-directionally coupled logistic maps together with the distribution of a single logistic map given by eq.~\eqref{eq:density:fs}.

Note that, as for the uni-directional setup, the second moment already diverge within the region of synchronization for logistic and tent maps whereas for Bernoulli maps the second moment diverges at the synchronization transition. With increasing coupling the effect of the noise becomes stronger and, hence, the susceptibility is not symmetric around $\alpha = 0.5$.

\begin{figure}
\centering
	\includegraphics[width=0.9\columnwidth]{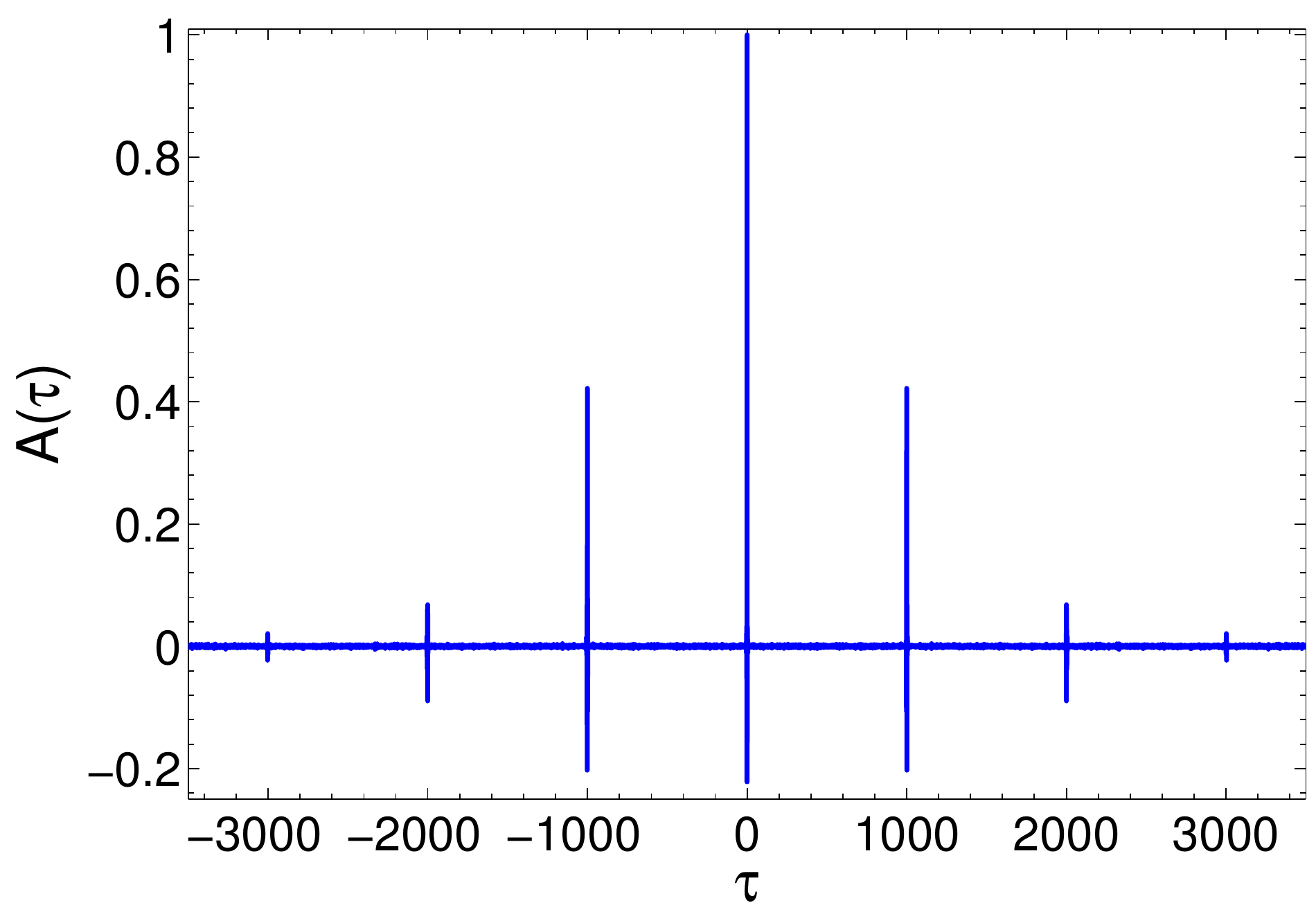}
\caption{Auto correlation $A(\tau)$ of $d_t$ for two uni-directionally coupled logistic maps with $a=4$, $\epsilon=0.7$, $\kappa=0$, $\tau=1000$ and $M=10^{-8}$.}
\label{fig:AutoCorrMoment}
\end{figure}

In the general case described by equation \eqref{eq:linA} and \eqref{eq:linB} for the uni and bi-directional setup, respectively, a time delay $\tau$ is present. The trajectory $d_t$ shows distinct auto-correlations at integer multiples of $\tau$ see Fig.~\ref{fig:AutoCorrMoment}. Unfortunately, we generally do not know the magnitude of the correlation between $d_t$ and $d_{t-\tau}$, which also changes for varying coupling parameters. Additionally the distribution of $f'$ is altered tremendously for a time delayed system compared to the distribution of a single unit, see \fig{\ref{fig:DistFDash}} (b). Hence we cannot calculate the second moment analytically but have to rely on numerical simulations.

\begin{figure}
\centering
	\includegraphics[width=0.75\columnwidth]{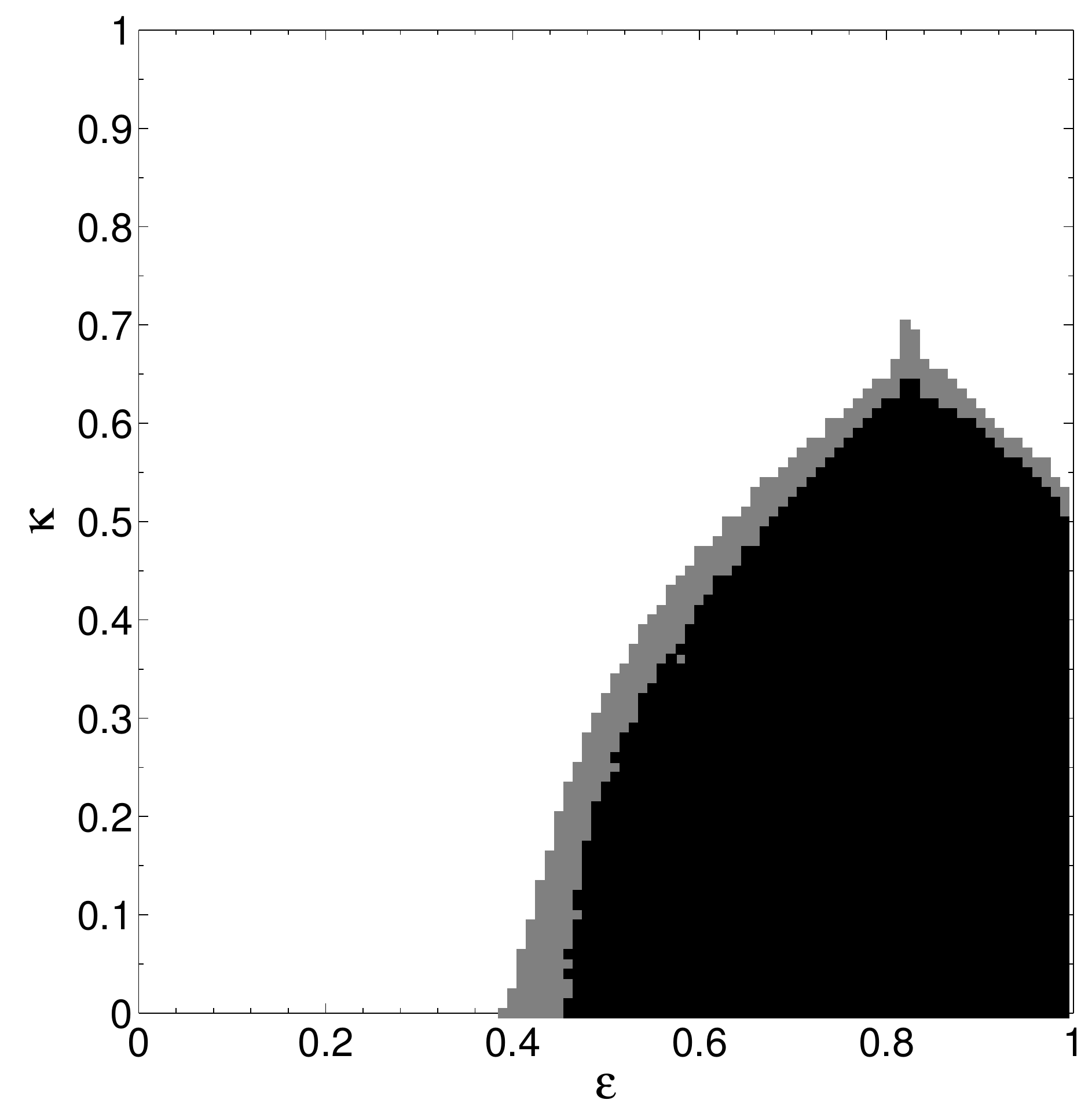}
	\caption{Phase diagram for two uni-directionally coupled logistic maps with $a=4$, $\tau=100$ and $M=10^{-8}$. 	
	Gray regime shows synchronization region, i.e., area where $C$ is larger than a threshold $\theta_c=0.999$. Black regime shows region where $\chi_2$ is finite (and the system is synchronized), i.e., $\chi_2$ smaller than a threshold $\theta_m=50$.}
	\label{fig:SynDivMoment}
\end{figure}

A numerical investigation of the time delayed system shows that the behavior of the moments is similar to a system with $\tau=0$.
For Bernoulli maps all moments diverge at the synchronization threshold whereas for logistic and tent maps $\chi_2$ already diverges inside the region of synchronization.
Numerical results for two uni-directionally coupled logistic maps are exemplary shown in \fig{\ref{fig:SynDivMoment}} as a binary plot where the system is assumed to be synchronized when the cross correlation exceeds some threshold $\theta_s$ and the second moment is assumed to be finite when it is smaller than some threshold $\theta_m$.

\subsection{Bit Error Rate}
\label{sec:ber}

The bit error rate (BER) measures the quality of the transmission of a (binary) message. For a proper reconstruction of the transmitted message it is crucial that the system is synchronized. However, the message is an external perturbation to the system that potentially destroys the synchronization. Hence the BER tells us how an external perturbation influences the synchronization, i.e., it is an indirect measure of the linear response.

In the following the sender transmits a binary message of the form $m_t = \pm M$ with $M \ll 1$ and $\mean{m} = 0$ which is reconstructed by the receiver by subtracting its own state from the received signal \cite{KinzelHandbook}. The recovered message $\tilde{m}_t$ is then given by
\begin{align}
	\tilde{m}_t = (x_t + m_t) - y_t = m_t - d_t \, .
\end{align} 
If both units were perfectly synchronized, i.e., $d_t = 0$, the original message would be perfectly recovered by the receiver. 
For a successful reconstruction of a binary message it is sufficient that both messages, $m_t$ and $\tilde{m}_t$, have the same sign, $m_t \tilde{m}_t > 0$.

The quality of the reconstruction is given by the BER $r$ which is defined as the the number of incorrectly recovered bits normalized by the total number of received bits,
\begin{align}
	r = \frac{\# \text{ incorrectly recovered bits}}{\# \text{ received bits}} \, .
\end{align} 
Note that by just guessing the bits we would be correct in half of the cases on average and hence obtain $r = 0.5$.

The BER is related to the distance $d$ by an integral over the distribution $\rho(d)$. If the absolute value of $d$ is less than the absolute value of the messages, $|d_t| < M$, the message is  recovered correctly with probability $1$ because the sign of the message is not changed. In the other cases, $|d_t| > M$, the sign of the message is changed with probability $\frac{1}{2}$, thus the BER is given by
\begin{align} \label{eq:ber}
\begin{split}
	r &= \frac{1}{2} \left( \int\limits_{-\infty}^{-M} \rho(d) \,\mathrm{d}d + \int\limits_{M}^{\infty} \rho(d) \,\mathrm{d}d \right) \\
	 &= \frac{1}{2} \left( 1 - \int\limits_{-M}^{M} \rho(d) \,\mathrm{d}d \right) \, .
\end{split}
\end{align} 
Note that this definition of the BER assumes that the system has relaxed to a stationary distribution $\rho(d)$. In fact, this definition may be only an upper bound since lower BERs may be achievable if one uses additional information about the transmitted signals \cite{EinDor:60:799}.

\begin{figure*}
\centering
	\includegraphics[width=0.32\textwidth]{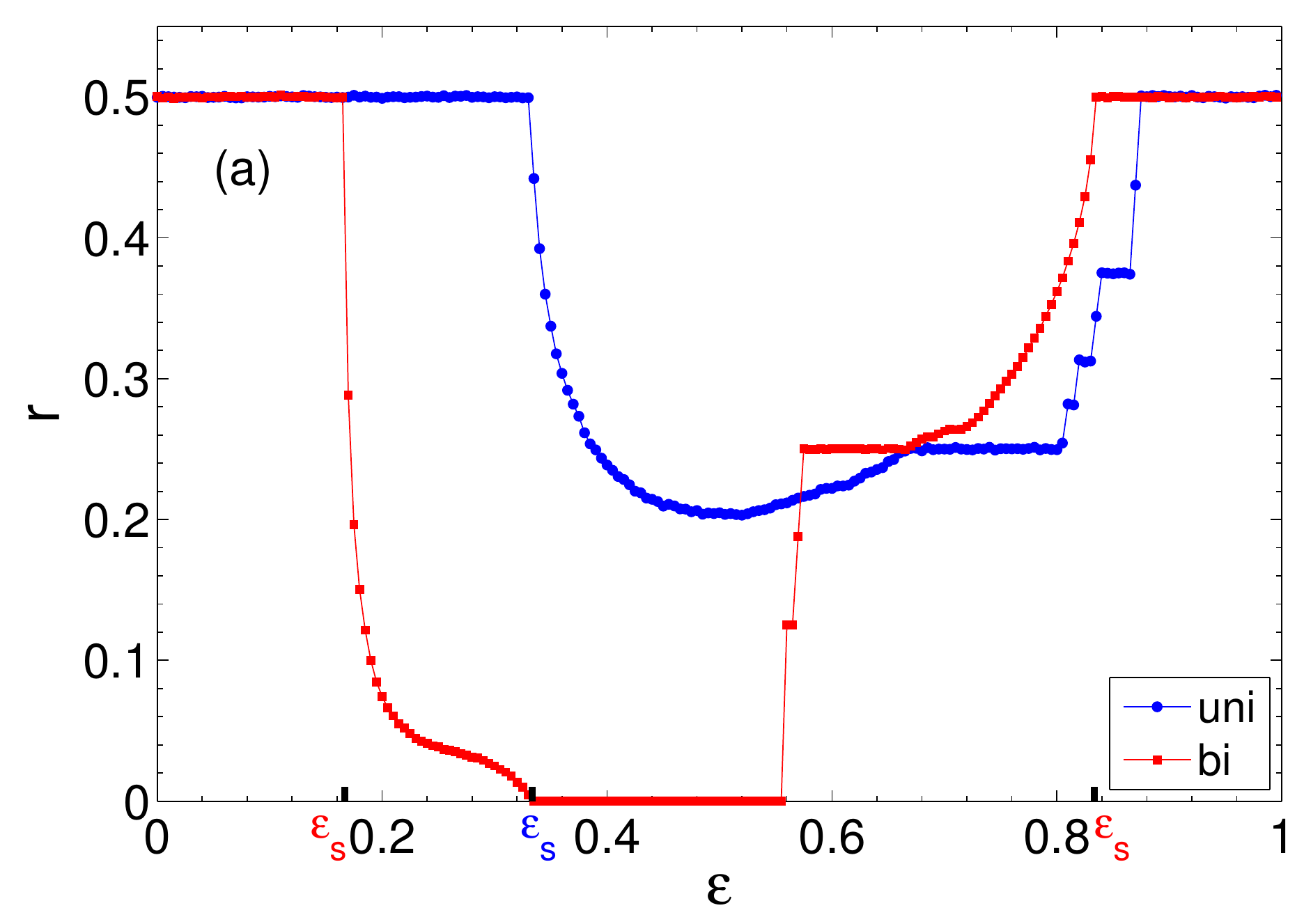}
	\includegraphics[width=0.32\textwidth]{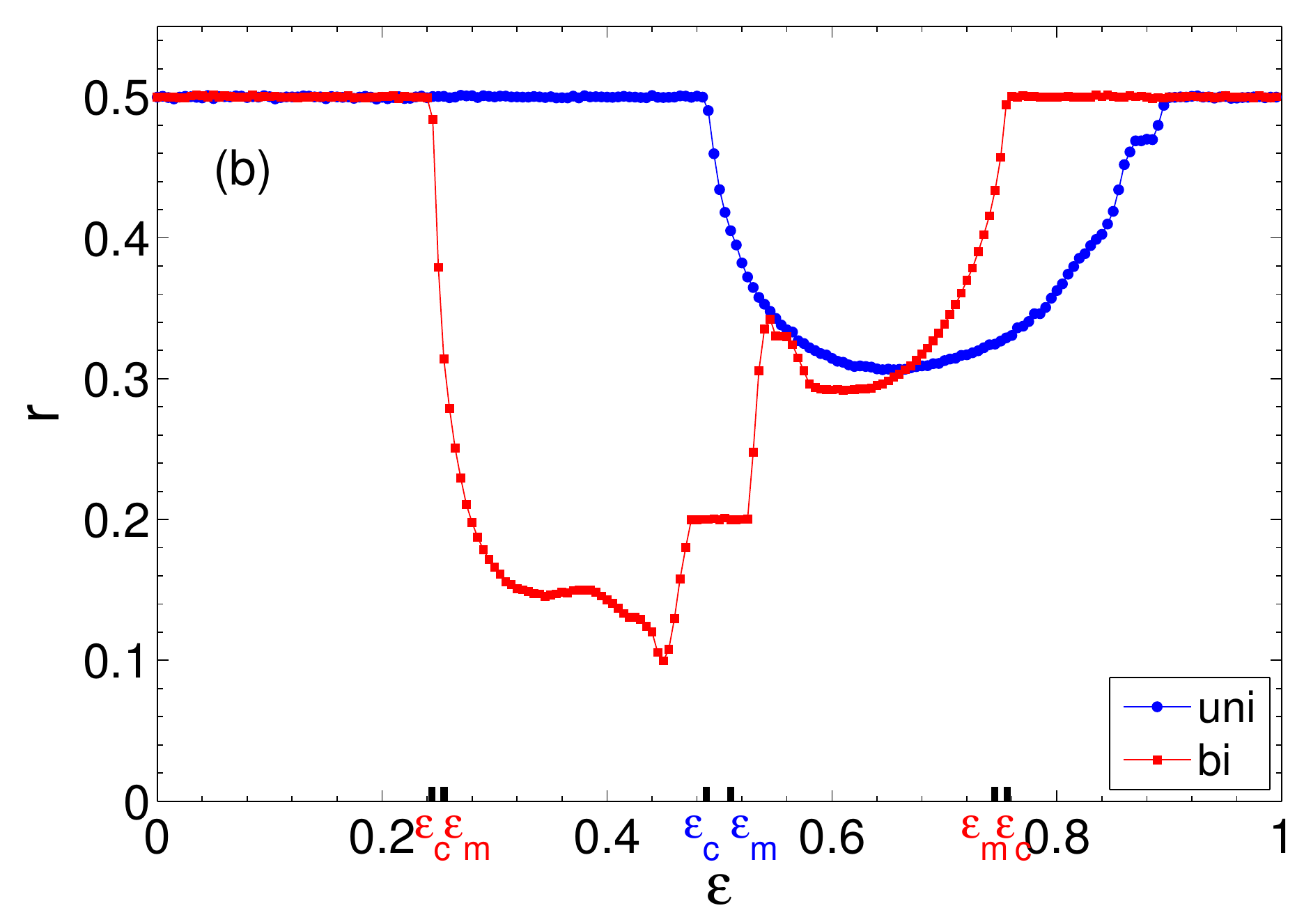}
	\includegraphics[width=0.32\textwidth]{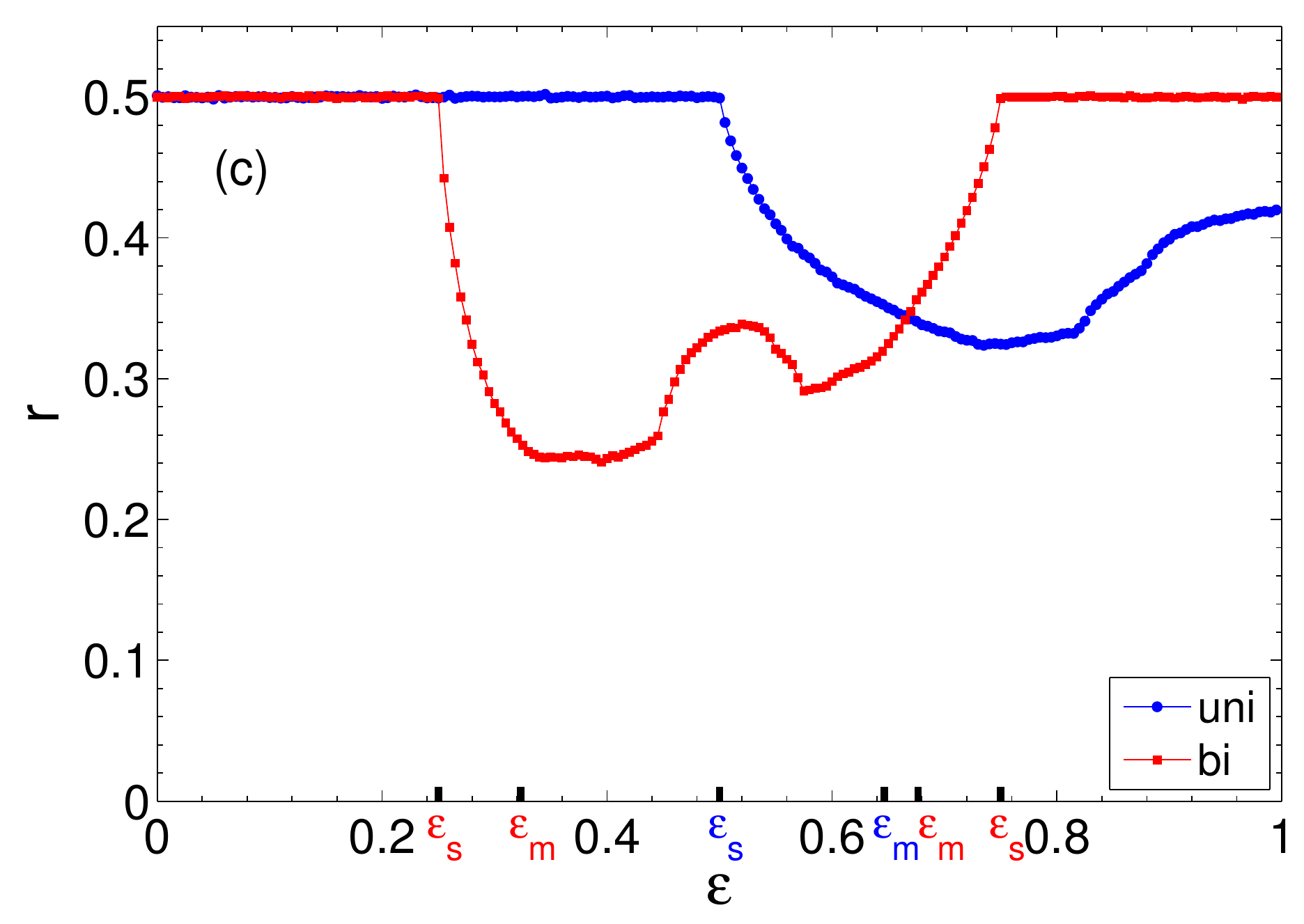}
	\caption{Bit error rate $r$ for uni- and bi-directionally coupled (a) Bernoulli (b) tent and (c) logistic maps with $a=1.5$, $a=0.4$ and $a=4$, respectively. Other parameters are $\tau = 0$, $\kappa = 0$ and $M=10^{-8}$. The synchronization transition is indicated by $\epsilon_s$ and the transition to a finite second moment by $\epsilon_m$.}
	\label{fig:BER_EpsScan}
\end{figure*}

The distribution $\rho(d)$ is known analytically only in some special cases, see Appendix. In general, the BER has to be determined by means of computer simulations. Figure \ref{fig:BER_EpsScan} shows results from simulations for the uni- and bi-directional setup without time delay for Bernoulli, tent and logistic maps. 
For an unsynchronized system the BER is at its maximum $r=0.5$. In contrast to the second moment which becomes finite at the synchronization transition only for Bernoulli maps, the BER drops down to smaller values for all maps as soon as the system synchronizes.
That means that the BER is smaller than $r=0.5$ although $d$ has large excursions from the synchronization manifold and the second moment diverges. Hence the sign of the deviation has correlations to the original message even though its amplitude has still a broad distribution. In this parameter range information can already be successfully transmitted. Although the BER is very high, one may apply methods from information theory to derive the message.

\begin{figure*}
\subfigure[Bernoulli map, uni-directional, $\epsilon = 0.4$]{
	\includegraphics[width=0.3\textwidth]{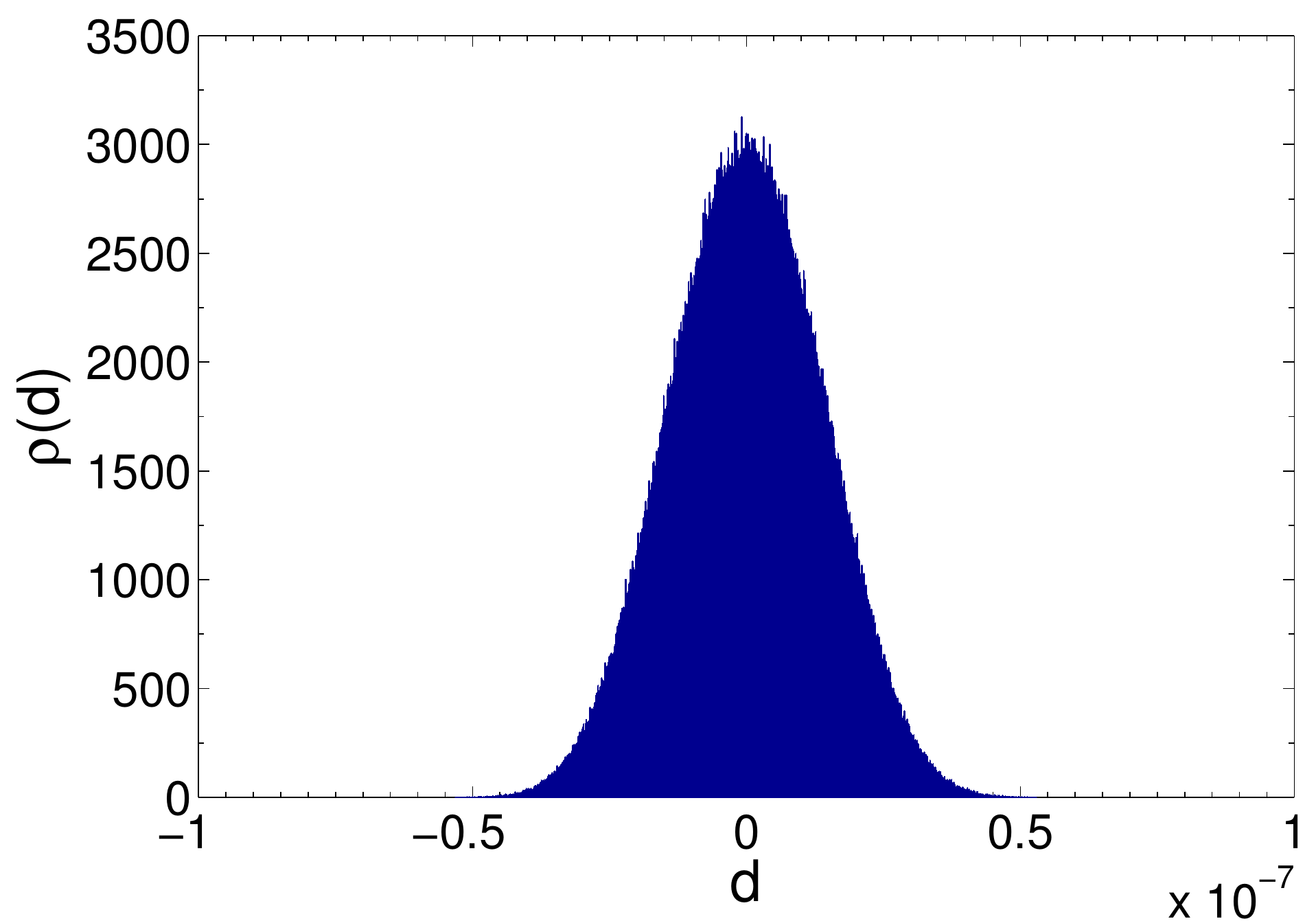}
}
\subfigure[Bernoulli map, uni-directional, $\epsilon = 0.6$]{
	\includegraphics[width=0.3\textwidth]{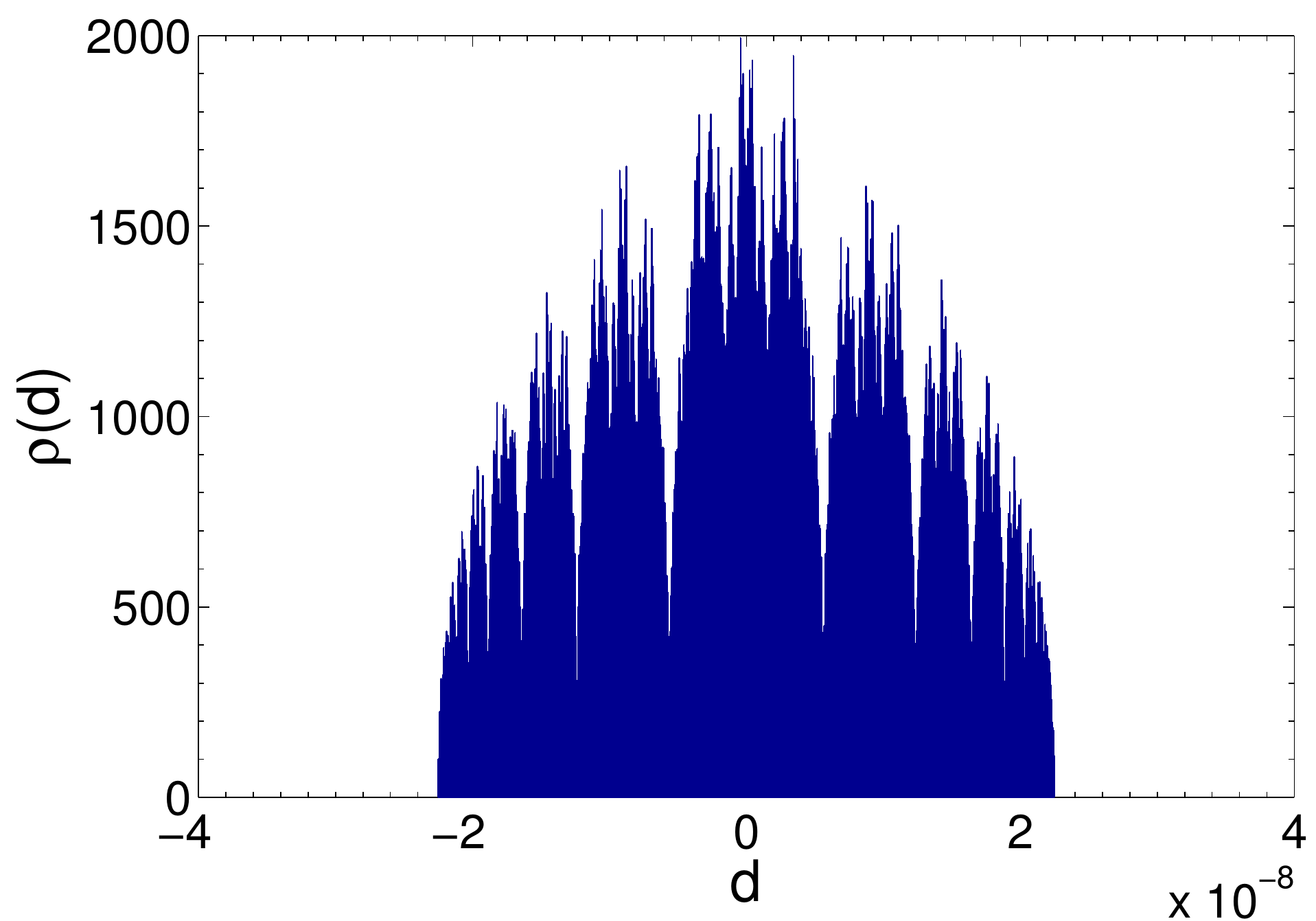}
}
\subfigure[Bernoulli map, uni-directional, $\epsilon = 0.8$]{
	\includegraphics[width=0.3\textwidth]{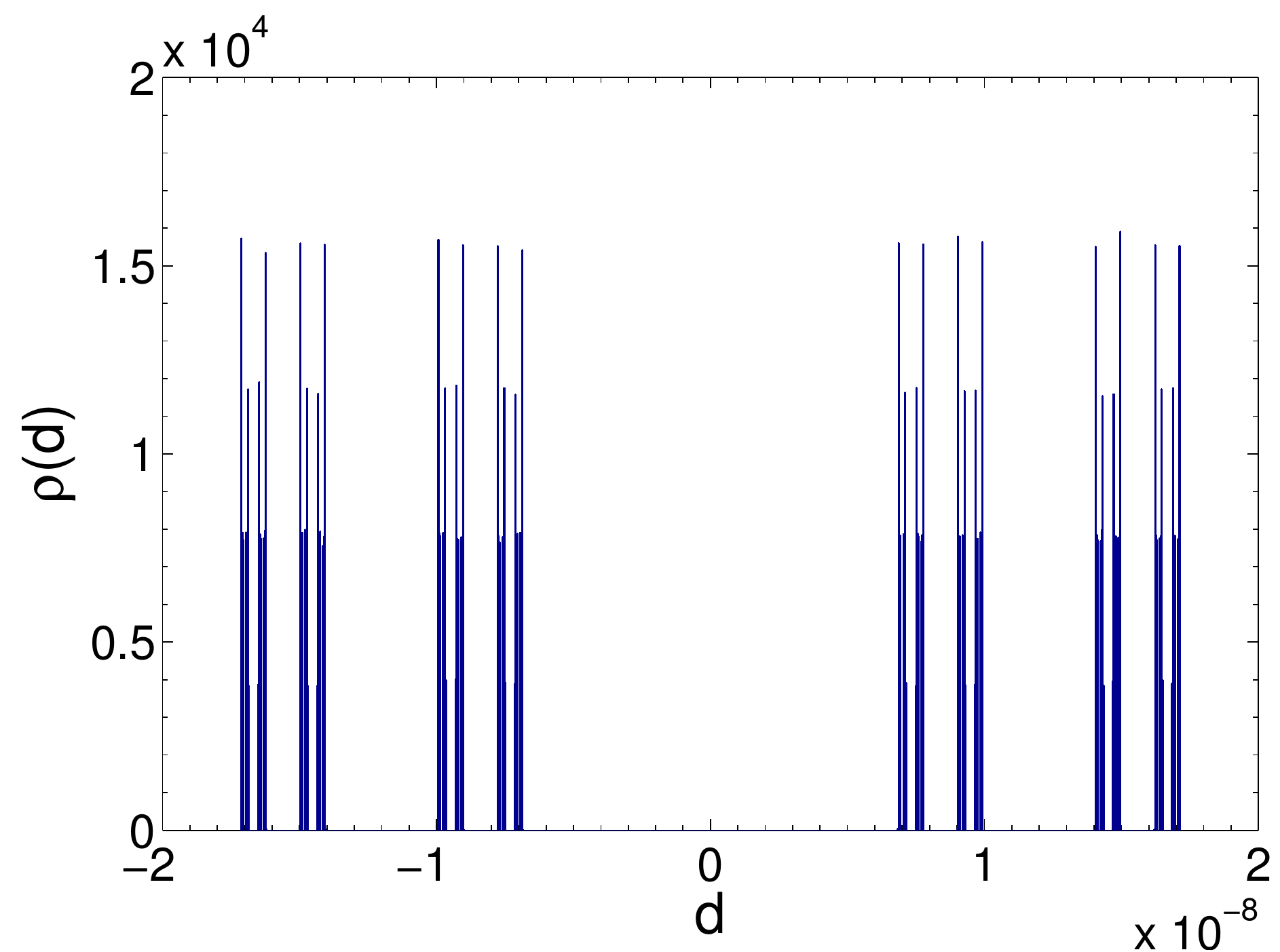}
}
\subfigure[Tent map, bi-directional, $\epsilon = 0.47$]{
	\includegraphics[width=0.3\textwidth]{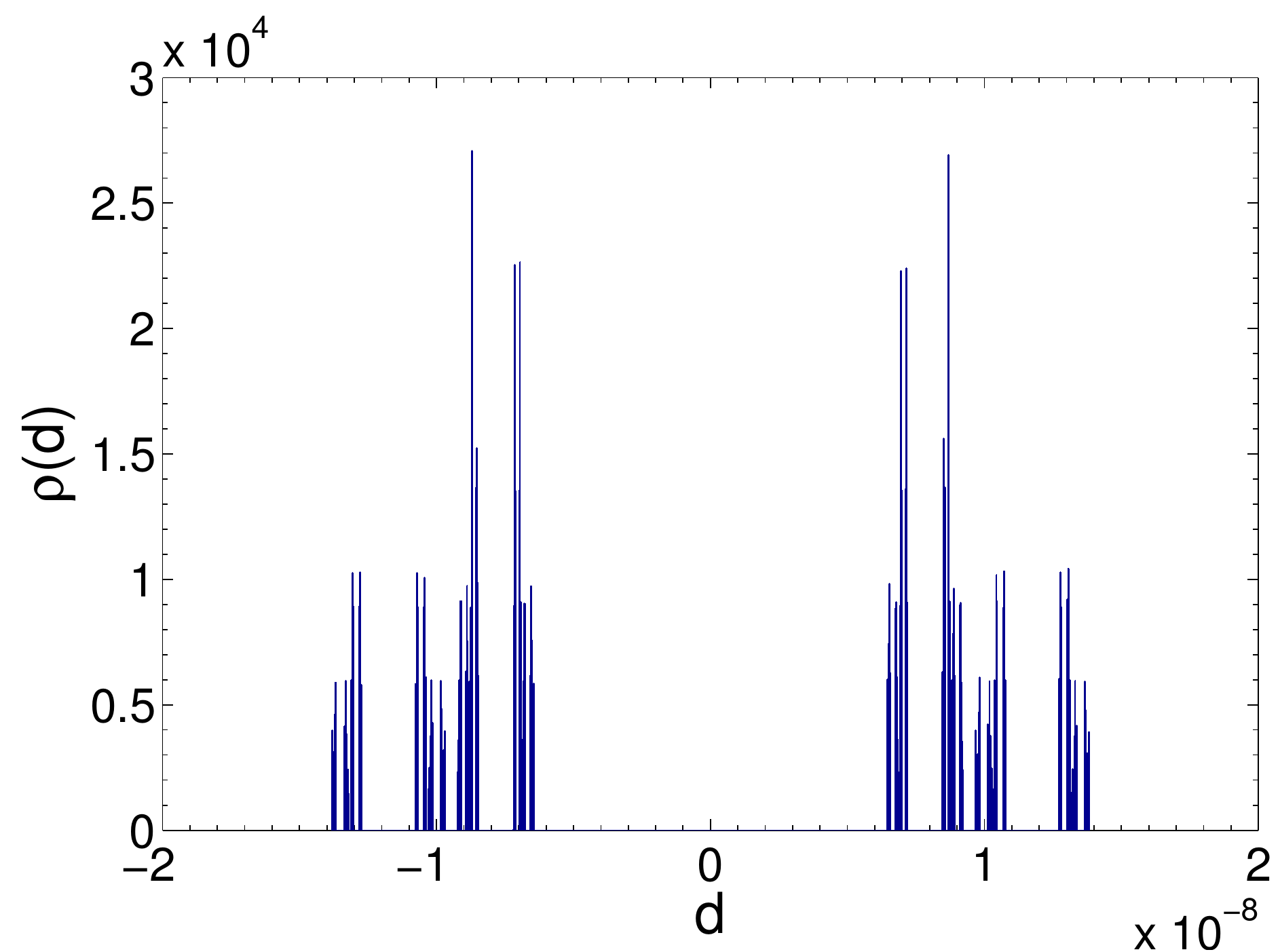}
}
\subfigure[Tent map, bi-directional, $\epsilon = 0.25$]{
	\includegraphics[width=0.3\textwidth]{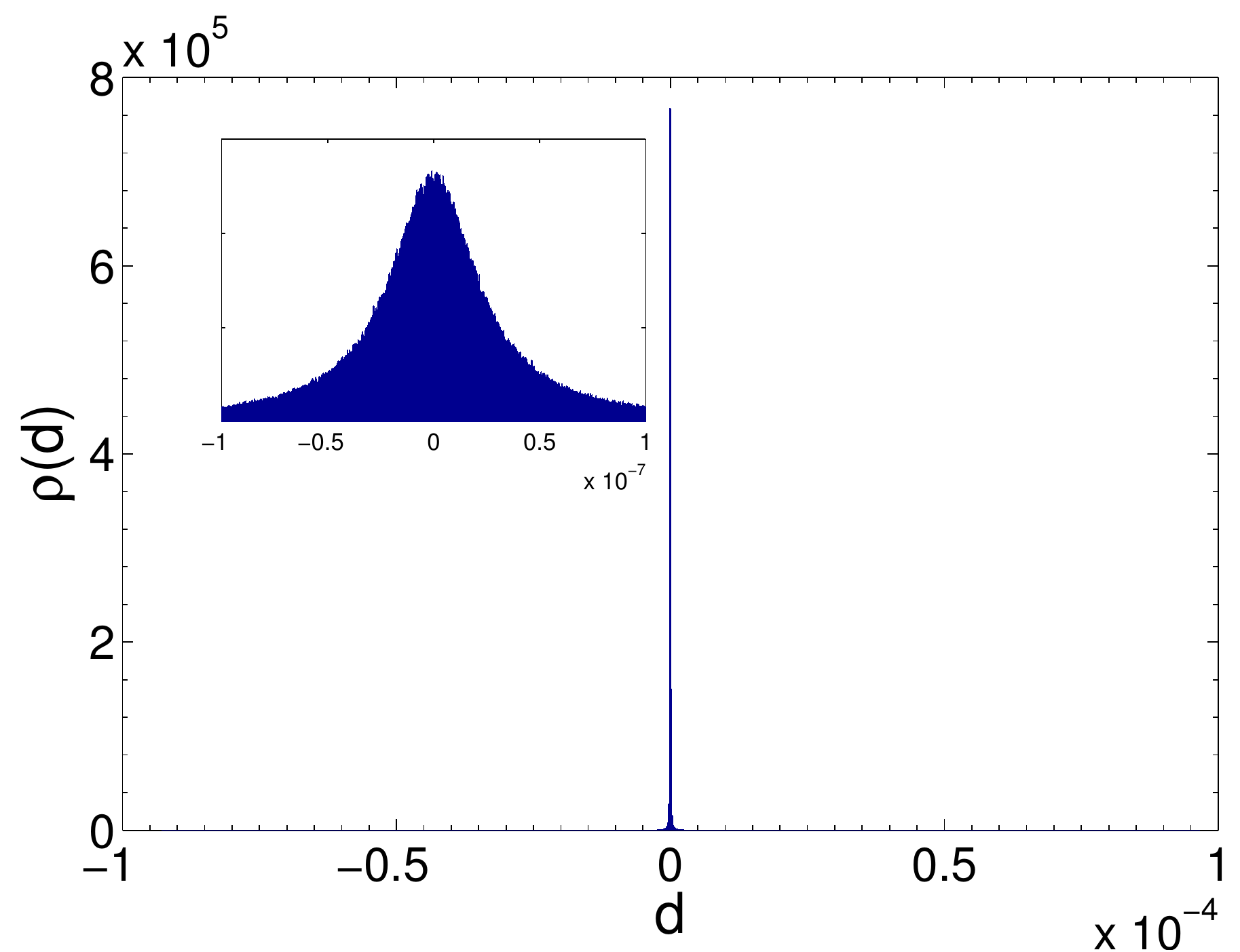}
}
\subfigure[Logistic map, bi-directional, $\epsilon = 0.8$]{
	\includegraphics[width=0.3\textwidth]{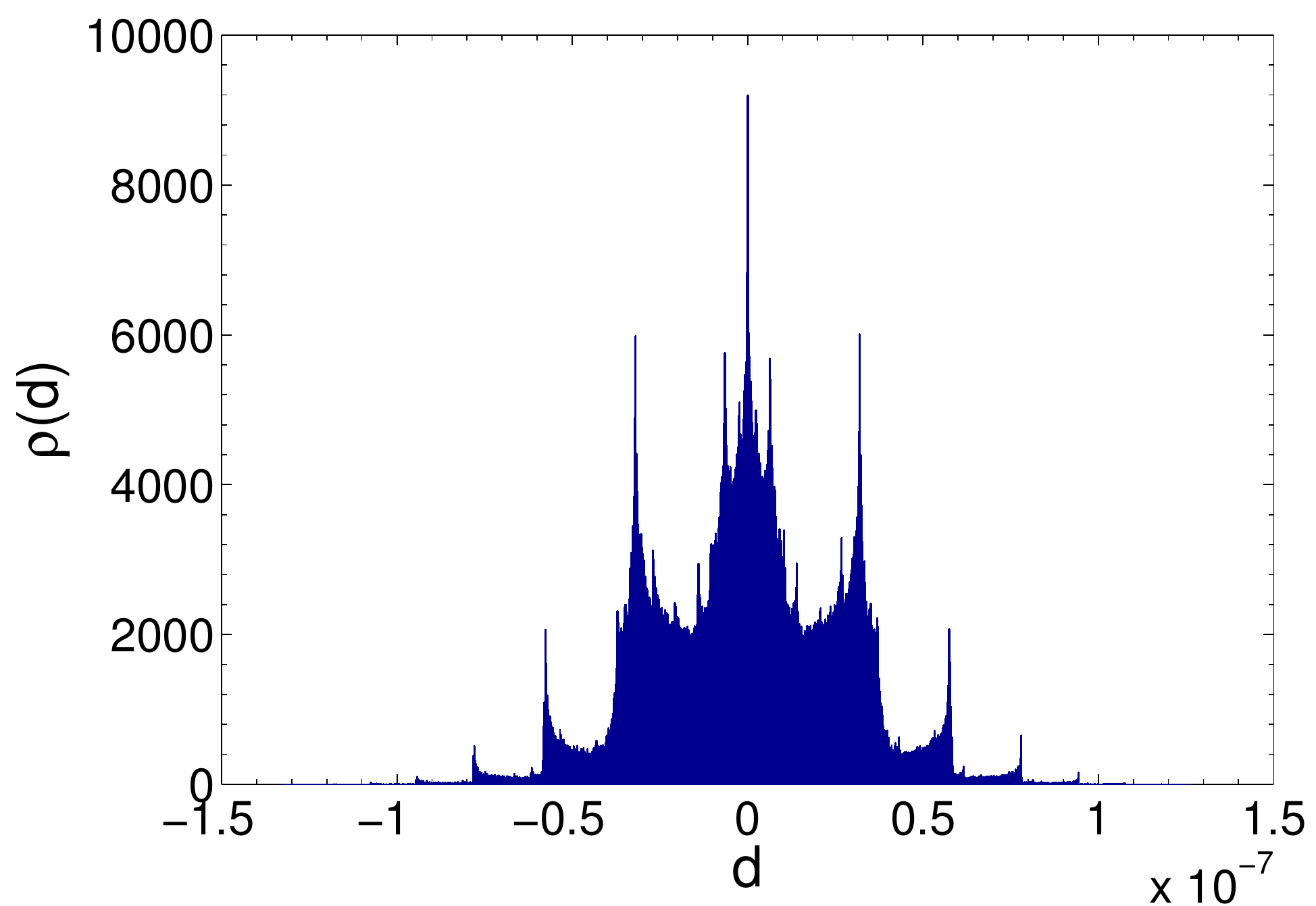}
}
\caption{Probability distribution $\rho(d)$ for different setups with parameters $\tau = 0$, $\kappa = 0$ and $M=10^{-8}$.}
\label{fig:histos}
\end{figure*}

The BER shows a staircase structure for Bernoulli and also for tent maps. As derived in the Appendix this devil's staircase is related to a fractal structure of the distribution $\rho(d)$.
For Bernoulli maps we find a broad distribution of $d$ for small $\epsilon$, see \fig{\ref{fig:histos}}(a). With increasing $\epsilon$ the distribution obtains more and more a fractal structure while the support stays connected, see \fig{\ref{fig:histos}}(b), until eventually the distribution changes to a peaked structure with a fractal support, see \fig{\ref{fig:histos}}(c). 
For the uni-directional setup the peaked distribution occurs for $\epsilon > \frac{2}{3}$ and the BER locks into rational values $r = \frac{k}{2^q}$ with $k$ and $q$ being natural numbers and $\frac{1}{4} \le r \le \frac{1}{2}$. For the bi-directional setup the distribution is peaked for $\frac{1}{3} \le \epsilon \le \frac{2}{3}$ and the BER is zero for $\frac{1}{3} \le \epsilon \le \frac{5}{9}$ since the absolute values of the distances are less than the message amplitude, $|d| < M$, in this interval.
Note that the fractal properties of the distribution $\rho(d)$ are related to the theory of iterated function systems \cite{Barnsley:Book}. For Bernoulli maps Eq.~(\ref{eq:dtEinfach}) and (\ref{eq:dtZweifach}) give iterations of two linear functions with $m_{t} = \pm 1$. Iterating a few randomly chosen functions can lead to fractal distribution.
For details see Appendix.

The distribution $\rho(d)$ for coupled tent maps has a similar behavior to the one for Bernoulli maps. It also shows a peaked structure which is related to a staircase in the BER, see \fig{\ref{fig:histos}}(d). But, in contrast to Bernoulli maps, the distribution for the tent map system can have very long tails, see \fig{\ref{fig:histos}}(e). These occur in the parameter range where the system is synchronized but the second moment diverges. For Bernoulli maps there is no such parameter range since both transitions take place at the same point.

For logistic maps the distribution $p(d)$ can also have power law tails. But unlike to Bernoulli and tent maps it does not show a peaked structure. The distribution has always a connected support due to the broad distribution of the multiplicative noise $f'$, see \fig{\ref{fig:histos}}(f).

The BER for a system with (large) time delay is exemplary shown for bi-directionally coupled tent maps in Figure \ref{fig:BER_EpsScan}. 
The synchronization transition is also indicated in the plot and illustrates that the BER, similar as for the undelayed system, is reduced as soon as the system synchronizes.

\begin{figure}
\centering
	\includegraphics[width=0.8\columnwidth]{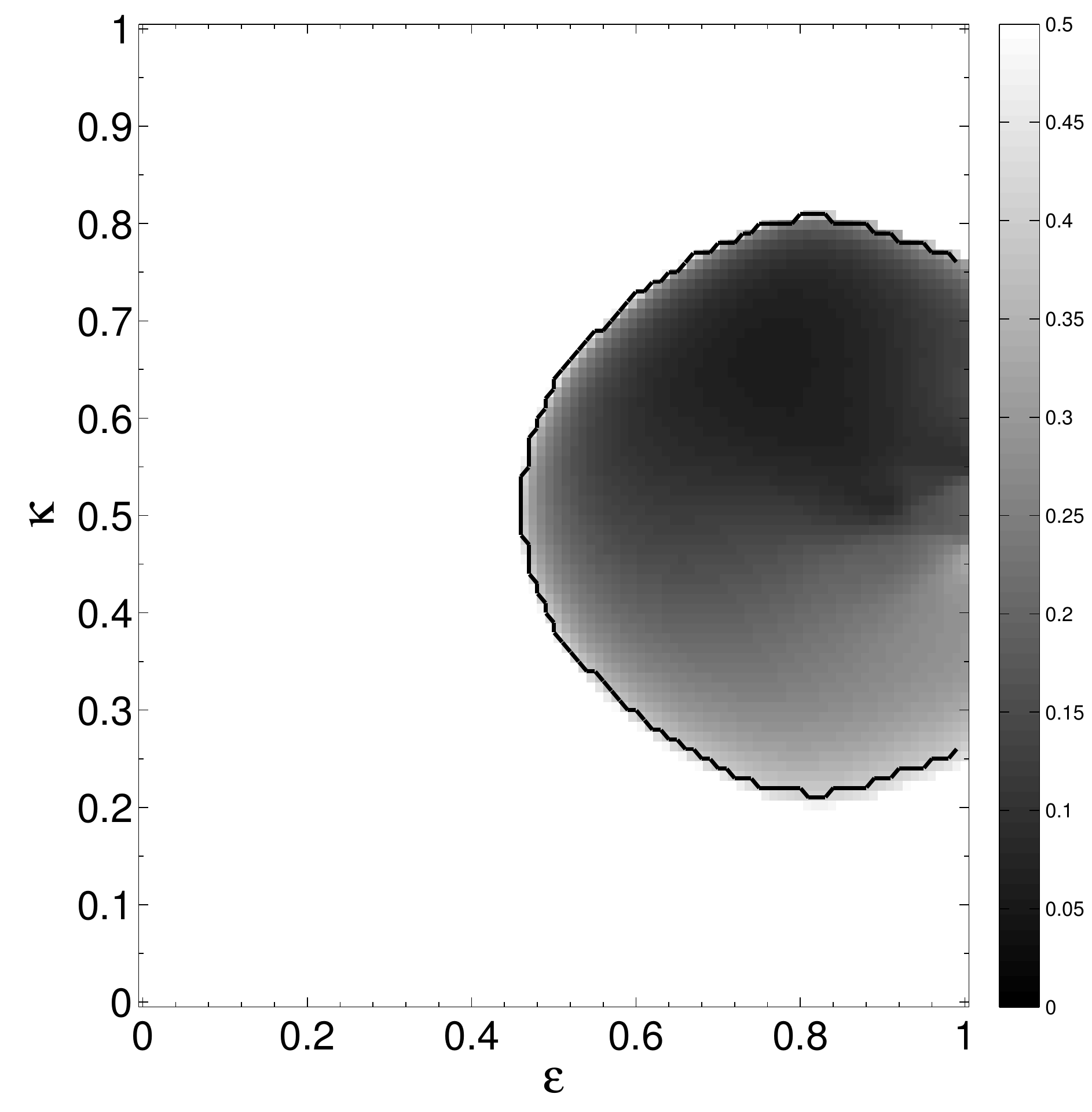}
	\caption{Bit error rate $r$ for a bi-directional tent map system with delay $\tau = 100$. The black solid line indicates the synchronization transition, i.e., the boundary for which all cross correlations are larger than a threshold, $C \geq \theta_s=0.999$. The other parameters are $a=0.4$ and $M=10^{-8}$.}
	\label{fig:BER_EpsKapScan}
\end{figure}

Note that our results show that the integral over the distribution of distances $\rho(d)$, \eq{\ref{eq:ber}}, which determines the BER, is insensitive to long tails of $\rho(d)$. Although the responses $\chi_n$ diverge at different coupling strengths $\epsilon_n$, the BER is smooth as a function of $\epsilon$.

In our model the BER is rather high compared to reported values of analog and other digital systems. Hence our investigation just gives a qualitative explanation of chaos pass filter. But note, that we are using bits of length $L=1$ for our investigation. The bit error can be exponentially reduced by increasing $L$.

\subsection{Resonances}

In the previous subsections we investigated the linear response of synchronized chaotic units to a random perturbation $m_t$. We in particular analyzed the distance $d$ of the units given by the linearized equations \eqref{eq:linA} and \eqref{eq:linB} for the uni and bi-directional setup, respectively. For Bernoulli maps these linear equations have constant coefficients and hence any arbitrary perturbation can be decomposed into its Fourier modes. In this special case it is sufficient to investigate the linear response of the system to a harmonic perturbation of the form $m_t = M \exp(-i \omega t)$. The system responds with the identical frequency $d_t = D \exp(-i \omega t)$, and the amplitude of the recovered signal $\tilde{m} = m_t - d_t$ is given by
\begin{equation}
	| \tilde{m} | = | M - D | \: ,
\end{equation} 
with the complex amplitude $D$. We obtain the following results
\begin{itemize}
	\item uni-directional case
		\begin{equation}
D = M \frac{\epsilon (1 - \kappa) a }{e^{- i \omega (\tau+1)} - (1 - \epsilon) a e^{- i \omega \tau} - \epsilon \kappa a}
\label{eq:uniMTilde}
		\end{equation} 
	\item bi-directional case
		\begin{equation}
D = M \frac{\epsilon (1 - \kappa) a }{e^{- i \omega (\tau+1)} - (1 - \epsilon) a e^{- i \omega \tau} - \epsilon (2\kappa - 1) a} \: .
		\end{equation} 
\end{itemize}

\begin{figure}
	\includegraphics[width=0.43\textwidth]{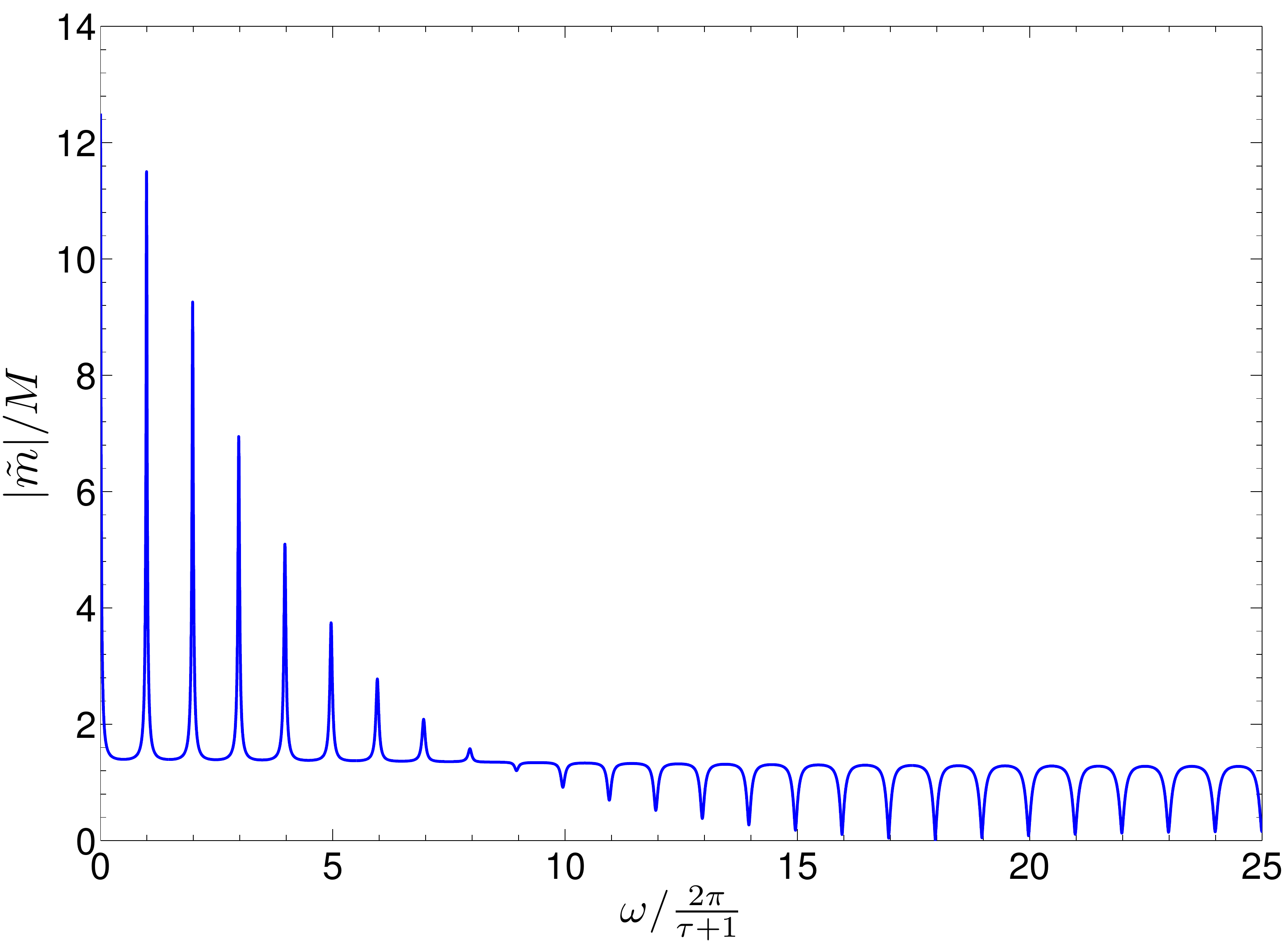}
	\caption{Amplitude of the reconstructed signal $\tilde{m}$ for different signal frequencies $\omega$. Bernoulli map, uni-directional coupling with $\epsilon = 0.8$, $\kappa = 0.55$ and $\tau = 50$.}
	\label{fig:reso}
\end{figure}

Figure~\ref{fig:reso} shows $| \tilde{m} |$ as a function of $\omega$ close to the phase boundary of synchronization. 
The amplitude of the reconstructed signal shows peaks (resonances) separated by a distance $2\pi/(\tau+1)$ which are caused by the first term of the denominator in equation \eqref{eq:uniMTilde}. The magnitude of the peaks is modulated due to the discrete nature of the systems equations. The resonances diverge at the synchronization boundary.

\begin{figure}
\centering
	\includegraphics[width=0.45\textwidth]{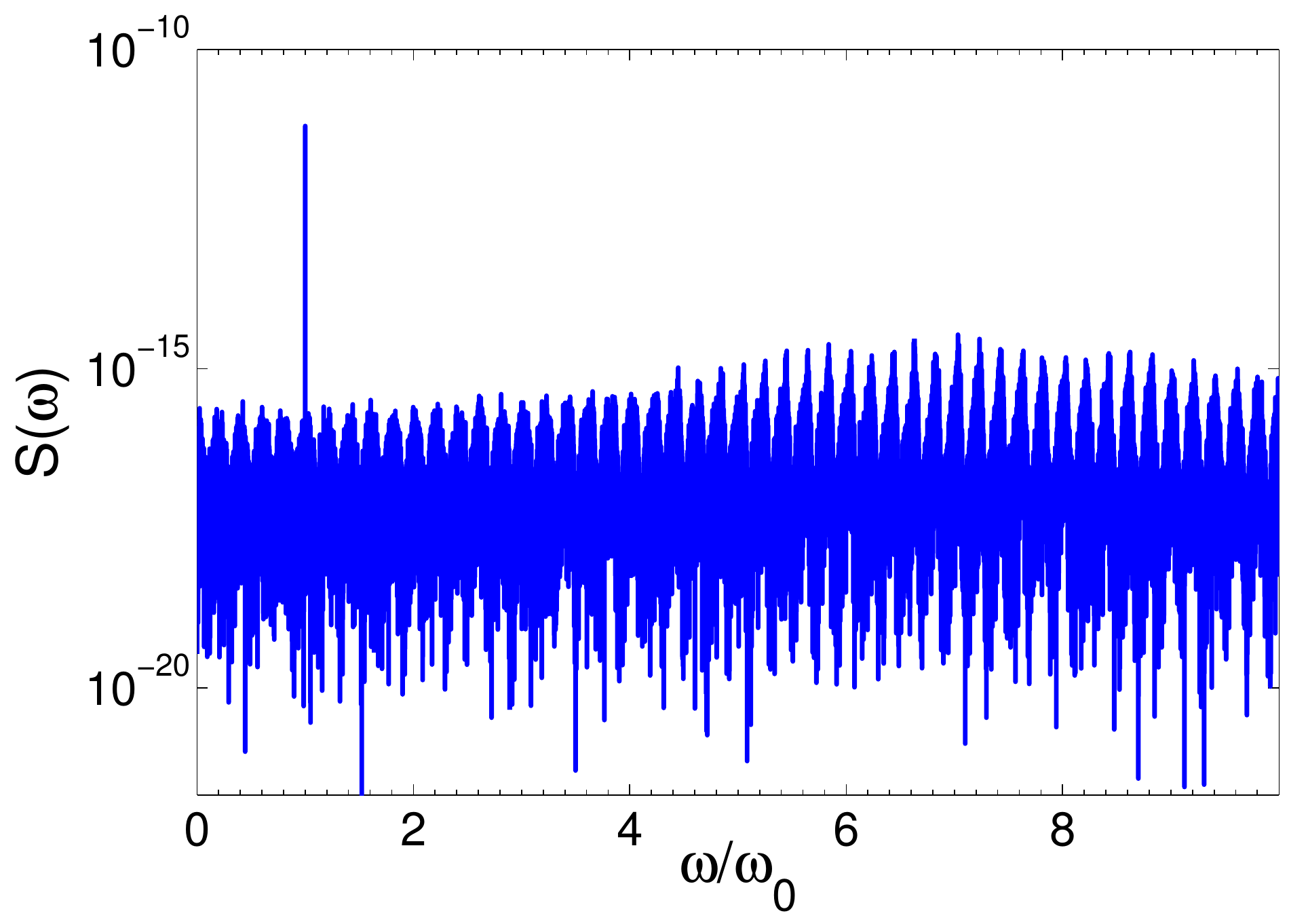}
	\caption{Power spectrum $S(\omega)$ of a reconstructed message $\tilde{m}$ in a semilogarithmic plot. Original message is of form $m_t= M \sin{\omega_0 t}$ with $M=10^{-8}$ and $\omega_0 =  5 \cdot 2 \pi / \tau$. Tent map, uni-directional coupling with $\epsilon = 0.6$, $\kappa = 0.4$ and $\tau = 100$. Note that the underlying periodic structure is due to the sampling frequency.}
	\label{fig:fourier}
\end{figure}

For the logistic and tent map the coefficients of the linear equations (\ref{eq:linA}) and (\ref{eq:linB}) depend on time. Thus an exact Fourier decomposition of the transmitted signal is not possible. However, our numerical results show that the corresponding Fourier component of the response $\tilde{m}_t$ shows resonances as well. The power spectrum of the recovered message has a clear peak at the frequency of the harmonic perturbation, see \fig{\ref{fig:fourier}}. Surprisingly, no higher harmonics are observed.

These results show that a chaotic system can function as a sharp harmonic filter. The transmitted signal is irregular and the harmonic perturbation is arbitrarily small. Nevertheless, the receiver can filter out this perturbation with high precision. It would be interesting to investigate this harmonic filter with respect to secret communication. In fact, the response of synchronized chaotic semiconductor lasers to a harmonic perturbation has been investigated in Ref.~\cite{Heil:2001}.

\subsection{Transverse Lyapunov Spectra}

For stable chaos synchronization the distance $d_t$ decays to zero such that the system's dynamics is restricted to the synchronization manifold. An external perturbation, however, drives the system away from this manifold. The competition between these two mechanisms results in the linear response investigated in the previous sections.

The relaxation to the synchronization manifold is described in terms of transverse Lyapunov exponents. For stable synchronization all transverse exponents are negative on average whereas in the unsynchronized case positive exponents exist. Close to the transition the maximum transverse exponent becomes very small until it eventually crosses zero at the transition. This results in a slowing down of the relaxation to synchronization and in a divergence of the response close to the transition. Diverging moments and power law tails are in particular related to local positive transverse Lyapunov exponents. Although the transverse exponent is negative on average in the synchronized case there can occur positive local ones which lead to temporary large excursions away from the synchronization manifold. These positive local transverse exponents may be caused by unstable periodic orbits or invariant subjects of the chaotic manifold \cite{Heagy:1995,Gauthier:1996}.

Following these arguments, one might expect that the qualitative behavior of the response functions $\chi_n$ or the bit error rate $r$ is related to the behavior of the largest transverse Lyapunov exponent $\lambda_{max}$. However, this is not true. For example, consider the simple driven system without delay, \eq{\ref{eq:dtEinfach}}. The transverse Lyapunov exponent is given by (\ref{eq:syncBed}),
\begin{equation}
	\lambda_{max} = \ln (1 - \epsilon) + \mean{ \ln |f'| } \,,
\end{equation} 
and decreases monotonically with increasing coupling strength $\epsilon$. However, the bit error rate $r$ first decreases with $\epsilon$ to a minimum value before it increases again, see \fig{\ref{fig:BER_EpsScan}}.

\begin{figure}
	\includegraphics[width=0.45\textwidth]{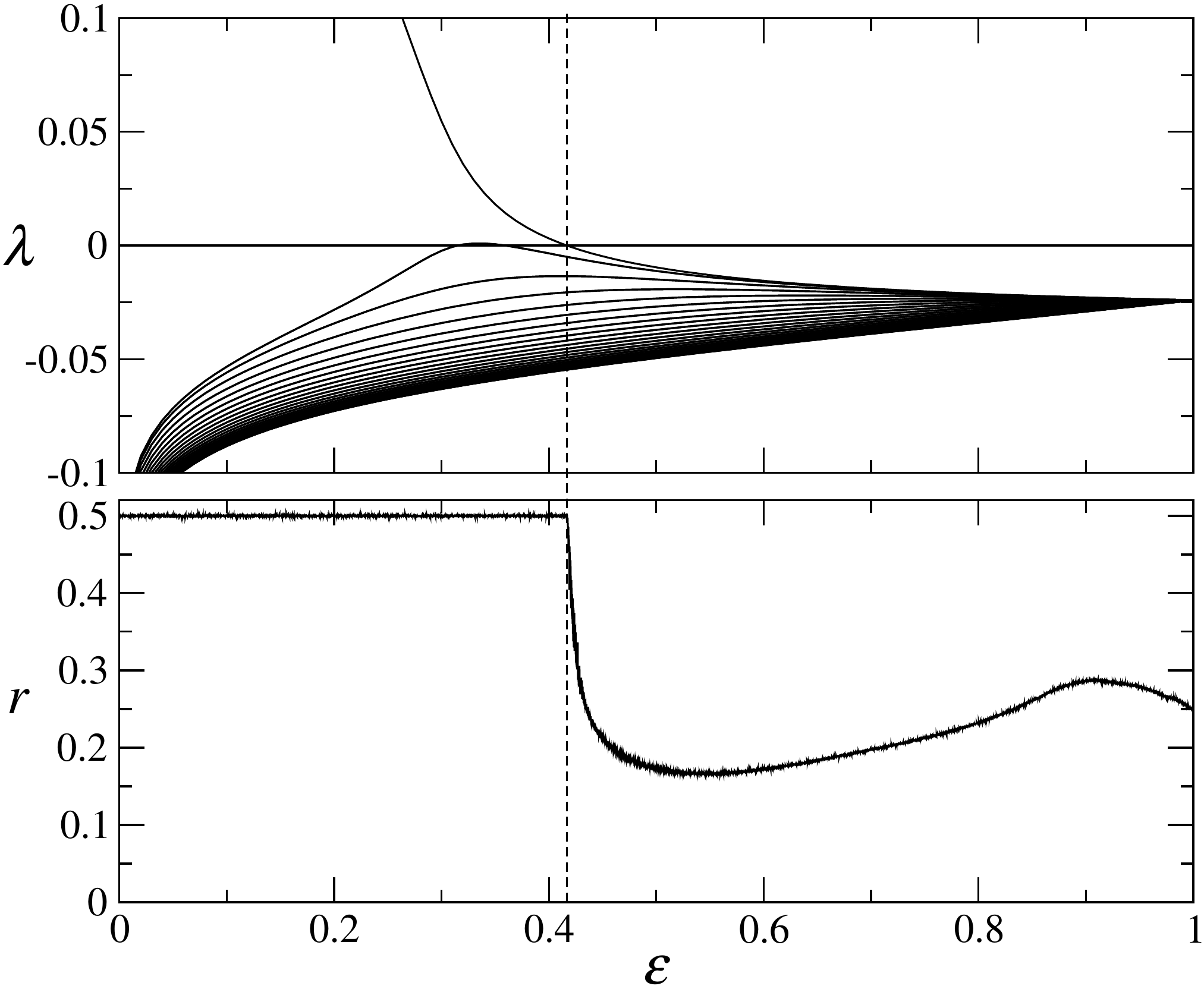}
	\caption{Spectrum of transverse Lyapunov exponents $\lambda$ and bit error rate $r$ for Bernoulli map, uni-directional coupling, $\kappa = 0.2$, $\tau = 50$.}
	\label{fig:lyap}
\end{figure}

For a Bernoulli system with delay $\tau$ the spectra of Lyapunov exponents can be calculated by solving polynomial equations of order $\tau$ \cite{Sublattice}. 
\fig{\ref{fig:lyap}} shows the transverse Lyapunov spectrum as a function of $\epsilon$ for the self-feedback strength $\kappa = 0.2$. Again, the qualitative behavior of the largest exponent is not related to the one of the bit error rate $r$. Except for the synchronization transition where the maximum exponent becomes negative and the BER becomes less than its maximum of $\frac{1}{2}$, the qualitative behavior of the Lyapunov exponents and the BER is not related.

\section{Small Networks}
\label{sec:ComplicatedModels}

Having discussed the linear response of a coupled two units system in the previous chapters, we are investigating the linear response of more complicated setups such as a chain of three units and a network of four units in this chapter. The motivation again stems from chaos communication. Hence, in all setups the external perturbation is added to the signal of the "sender" and can be thought of as a secret message.

The systems in this section are generally to complicated for an analytical discussion of the moments $\chi_n$ or the BER $r$. Hence we mainly rely on numerical simulations in the following.  

We define the second moment for different units by $\chi_{ij}$ where the index stands for the combination of unit $i$ and $j$, i.e., the distance $d_{ij} = x^j_t-x^i_t$ is used when evaluating Eqn.~\eqref{eq:chi}. Similar, for the BER $r_{ij}$ the signals of units $i$ and $j$ are compared.

\subsection{Chain of Three Units}

\begin{figure}
\centering
 	\includegraphics[width=0.8\columnwidth]{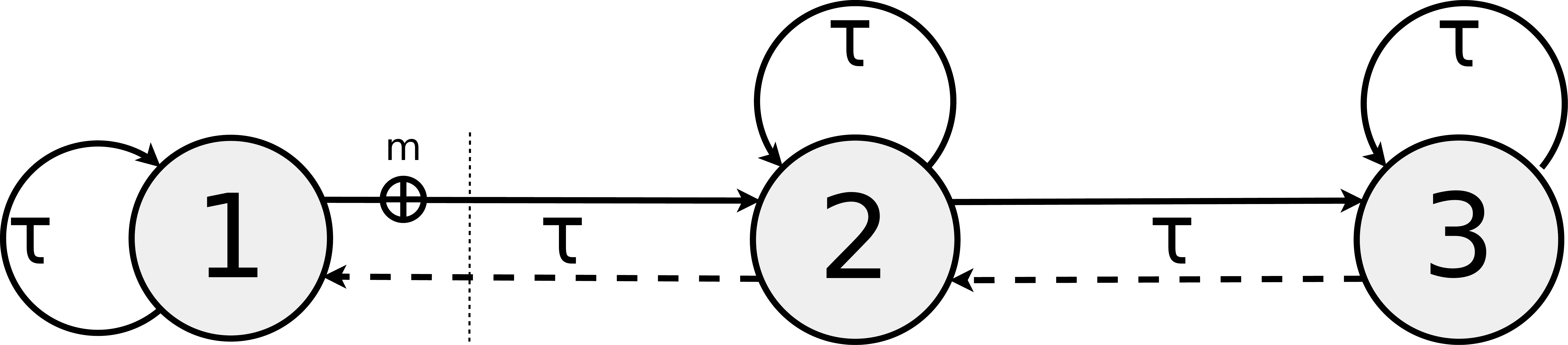}
	\caption{Setup of three coupled chaotic units with either uni- or bi-directional (dashed line) coupling. A perturbation $m$ is added to the exchanged signal at unit 1, i.e., the sender. The transmitted signal has a time delay $\tau$.}
	\label{fig:Setup3Chain}
\end{figure}

In the following the system comprises of three chaotic units which are arranged in a line such that the first unit is coupled to the second and the second to the third. The coupling can either be uni- or bi-directional and has a time delay. Optionally the units are subjected to a self-feedback with the same delay time. The setup is depicted in Fig.~\ref{fig:Setup3Chain} and the system's equations are 
\begin{itemize}
\item uni-directional setup
\begin{align}
	x_{t+1}^1 &= (1-\epsilon) f(x_t^1) + \epsilon f(x_{t-\tau}^1) \notag \\
	x_{t+1}^2 &= (1-\epsilon) f(x_t^2) + \epsilon \kappa f(x_{t-\tau}^2) \notag \\
	& \quad \quad + \epsilon (1-\kappa) \, f(x_{t-\tau}^1 + m_{t-\tau}) \notag \\
	x_{t+1}^3 &= (1-\epsilon) f(x_t^3) + \epsilon \kappa f(x_{t-\tau}^3) + \epsilon (1-\kappa) f(x_{t-\tau}^2) 
	\label{Eqn:3ChainSysUni}
\end{align}
\item bi-directional setup
\begin{align}
	x_{t+1}^1 &= (1-\epsilon) f(x_t^1) + \epsilon \kappa f(x_{t-\tau}^1) + \epsilon (1-\kappa) f(x_{t-\tau}^2)  \notag \\
	x_{t+1}^2 &= (1-\epsilon) f(x_t^2) + \epsilon \kappa f(x_{t-\tau}^2) \notag \\
	& \quad \quad + \epsilon (1-\kappa) \left( 1/2 \, f(x_{t-\tau}^1 + m_{t-\tau})  + 1/2 \, f(x_{t-\tau}^3) \right)  \notag \\
	x_{t+1}^3 &= (1-\epsilon) f(x_t^3) + \epsilon \kappa f(x_{t-\tau}^3) + \epsilon (1-\kappa) f(x_{t-\tau}^2) 
	\label{Eqn:3ChainSysBi}
\end{align}
\end{itemize}
with the coupling parameters $\epsilon$ and $\kappa$ as before.

For Bernoulli units without noise, $m_t = 0$, the synchronization properties can be analyzed analytically by a master stability function method \cite{Pecora-1998, Englert-2011}. In the limit of large delays one finds that for the uni-directional setup only complete synchronization can occur whereas for the bi-directional setup complete and sub-lattice synchronization exists \cite{Patterns}. 
For the uni-directional setup the system synchronizes completely in the parameter regions II + III of \fig{\ref{fig:SyncArea}} which is identical to the synchronization region of two uni-directionally coupled units. The bi-directional system synchronizes completely in region II whereas in region III only unit 1 and unit 3 are synchronized.

\begin{figure}
\centering
 	\includegraphics[width=0.45\textwidth]{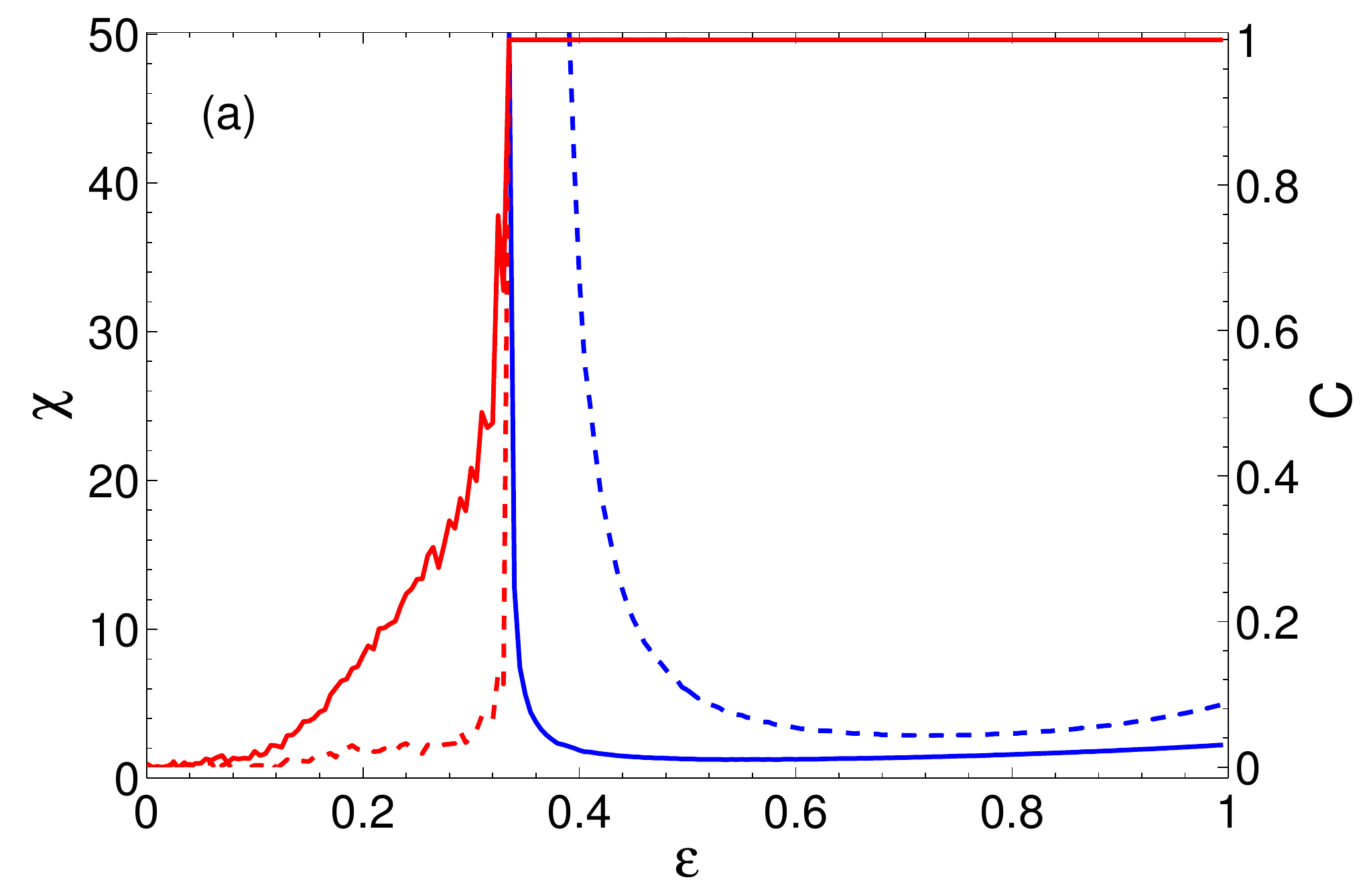}
 	\includegraphics[width=0.45\textwidth]{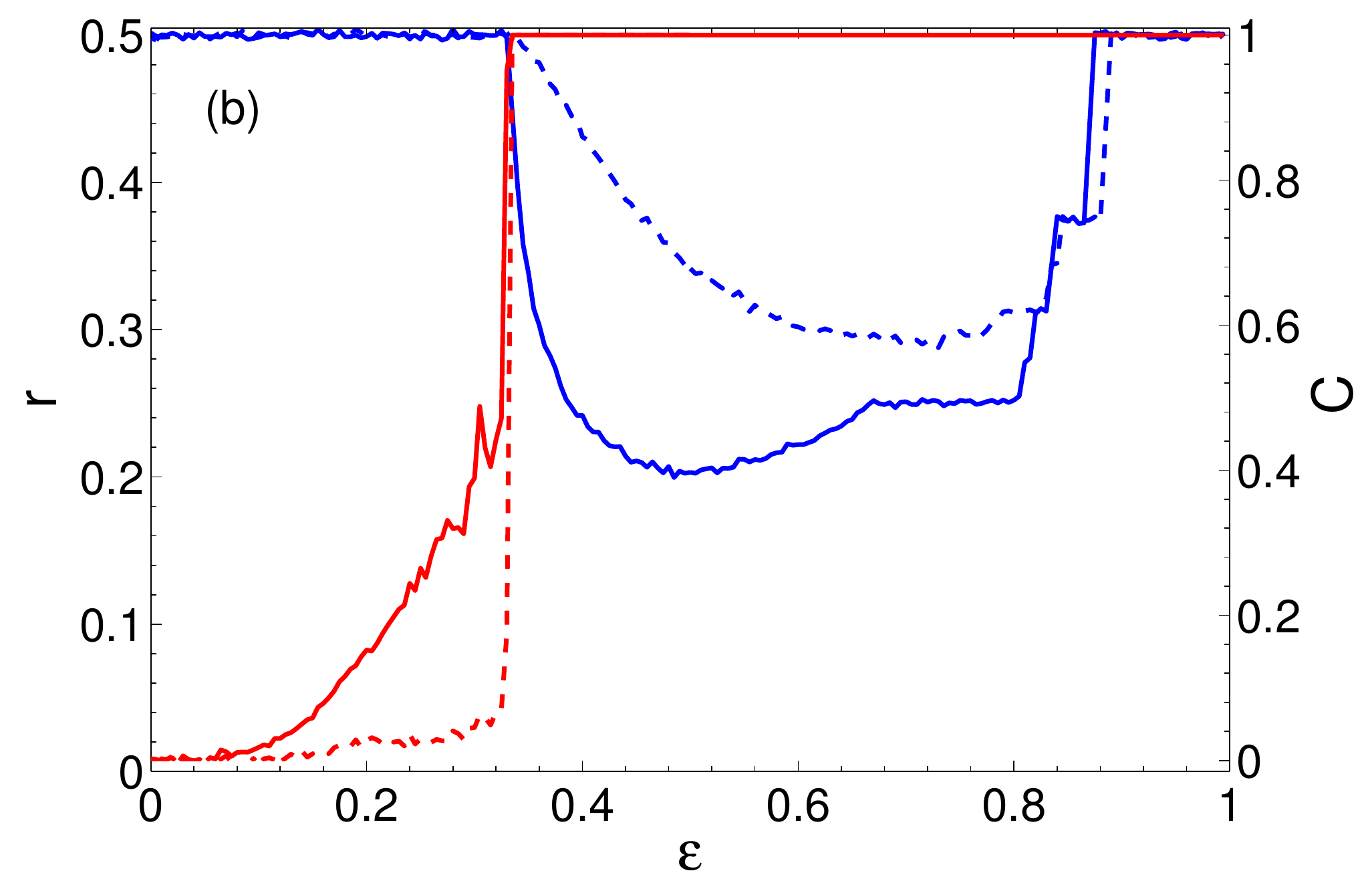}
	\caption{Panel (a) shows second moment $\chi_{ij}$ (blue curves) and cross correlation $C_{ij}$ (red curves) for uniformly distributed random noise, panel (b) shows bit error rate $r_{ij}$ (blue curves) and cross correlation $C_{ij}$ (red curves) for binary random noise as a function of $\epsilon$. Solid (dashed) line shows the respective results for the combination of units $ij=12$ ($ij$=13). Chain of uni-directionally coupled Bernoulli maps with parameters $a=1.5$, $\kappa=0$, $\tau=100$ and $M=10^{-8}$.}
	\label{fig:3ChainUniResults}
\end{figure}

In case of uni-directional coupling unit 1 imposes its chaotic behavior onto unit 2 which in turn imposes its behavior onto unit 3. All three units become synchronized for the same coupling parameters, however in the unsynchronized case the cross correlations are highest between unit 1 and 2. For a Bernoulli system, numerical results for the second moment and the BER together with the respective cross correlation for uniformly distributed and binary noise are shown in Fig.~\ref{fig:3ChainUniResults}. In contrast to the two units setup, where the second moment for Bernoulli maps diverge at the synchronization transition, the second moment $\chi_{13}$ diverges already inside the synchronized region. As we expect, the response is smallest for the combination of unit 1 and 2. The BER decrease for both combinations, $r_{12}$ and $r_{13}$, at the synchronization transition, similar to the two units setup, and is in general lowest for $r_{12}$. A devil's staircase occurs as well at which $r_{13}$ becomes the lower BER.

For the bi-directional setup unit 1 and 3 synchronizes perfectly, hence the distance $d_{13}$ and also the second moment $\chi_{13}$ is zero. A perturbation of the signal which unit 1 sends to unit 2 is completely filtered out and does not affect unit 3. This can easily be seen in the system's equations \eqref{Eqn:3ChainSysBi}. Unit 1 and 3 obtain the same external input from unit 2 and since the units are identical with the same internal dynamics they synchronize perfectly. Note that unit 2 does not necessarily need to be synchronized in order for perfect synchronization between unit 1 and 3 to occur. It acts as a relay which transmits the signals between the two units such that they can synchronize. Comparing the outgoing signal of unit 1, i.e. its internal dynamics plus the message $m$, with the outgoing signal of unit 3, the message can be recovered perfectly without any errors, thus the BER is $r_{13}=0$. 

\begin{figure}
\centering
 	\includegraphics[width=0.45\textwidth]{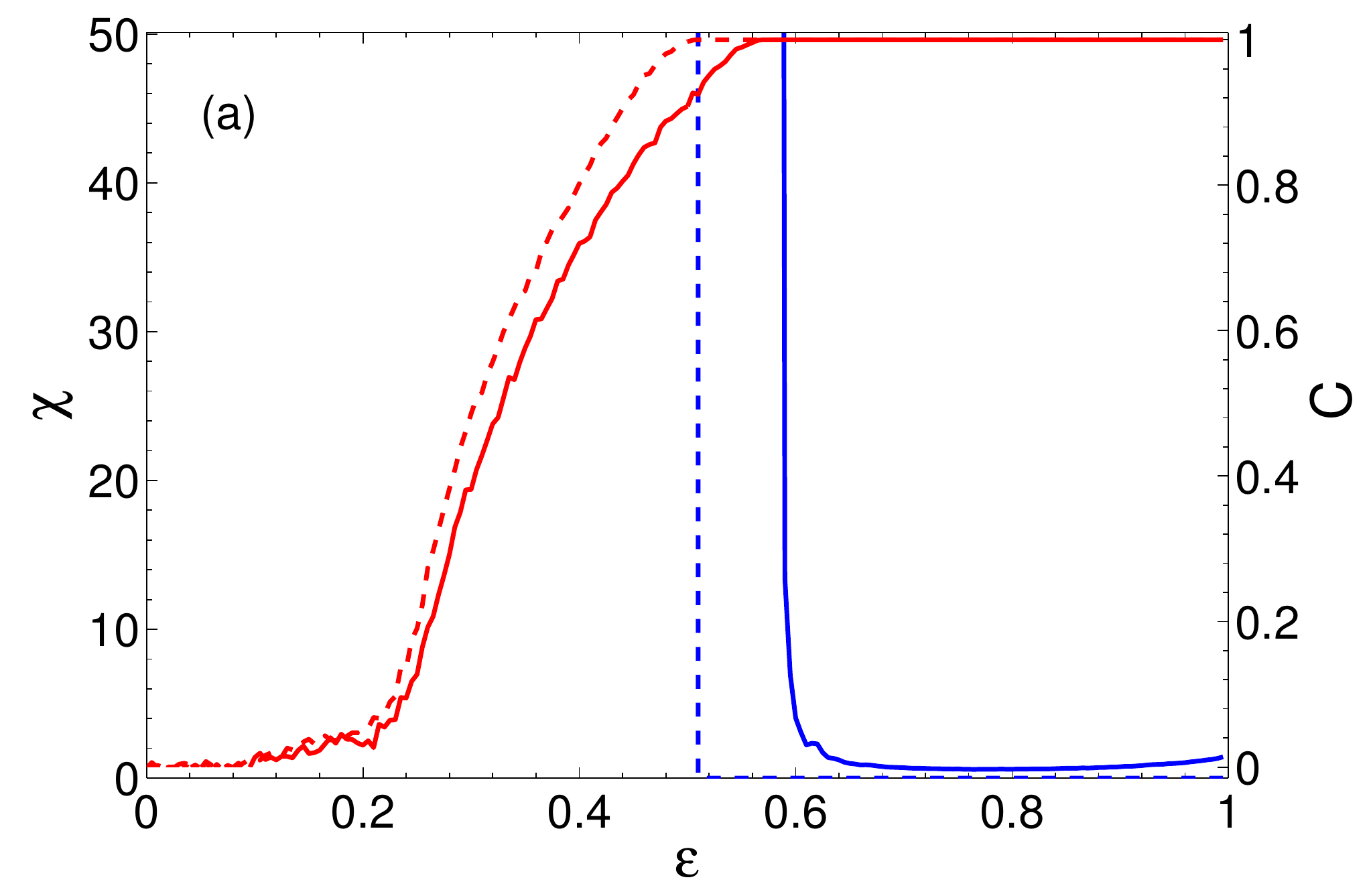}
 	\includegraphics[width=0.45\textwidth]{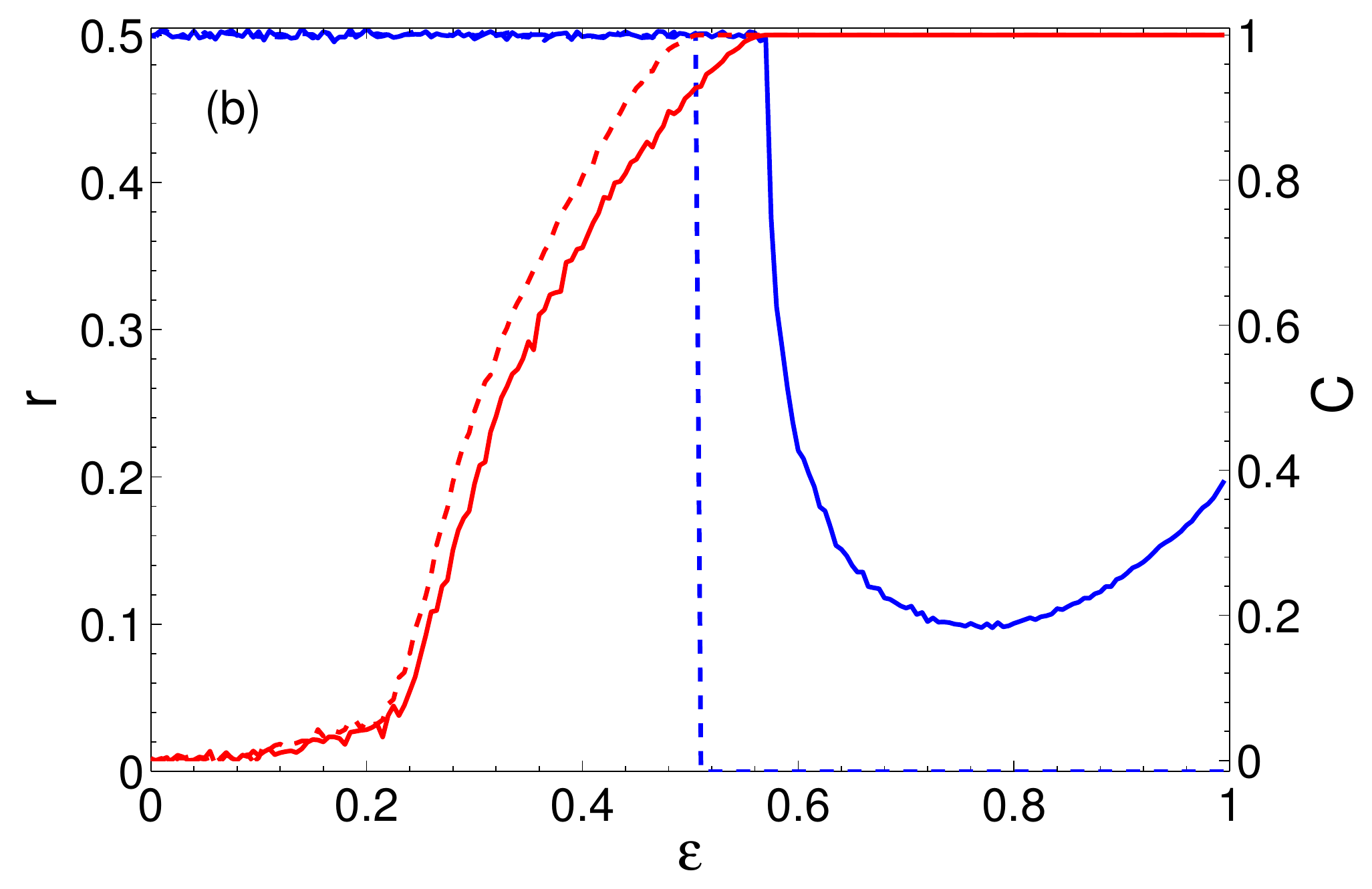}
	\caption{Panel (a) shows second moment $\chi_{ij}$ (blue curves) and cross correlation $C_{ij}$ (red curves) for uniformly distributed random noise, panel (b) shows bit error rate $r_{ij}$ (blue curves) and cross correlation $C_{ij}$ (red curves) for binary random noise as a function of $\epsilon$. Solid (dashed) line shows the respective results for the combination of units $ij=12$ ($ij$=13). Chain of bi-directionally coupled tent maps with parameters $a=0.4$, $\kappa=0.3$, $\tau=100$ and $M=10^{-8}$.}
	\label{fig:3ChainBiResults}
\end{figure}

Fig.~\ref{fig:3ChainBiResults} shows the second moment and the BER together with the cross correlations for a system of tent maps as a function of the coupling $\epsilon$. The coupling constant $\kappa$ was chosen such that sub-lattice synchronization occurs for some values of $\epsilon$. Sub-lattice synchronization exists in the region where $C_{13}=1$ but $C_{12}, C_{23} < 1$.  At the transition to sub-lattice synchronization $\chi_{13}$ and $r_{13}$ becomes zero.
At the transition to complete synchronization the BER $r_{12}$ decreases whereas the second moment $\chi_{12}$ is still diverging close to the boundary. But neither the BER $r_{12}$ nor the second moment $\chi_{12}$ becomes zero. The external noise prevents the trajectories, $x_1=x_3$ and $x_2$, of perfectly synchronizing. It causes the trajectories to deviate by a factor of the noise strength.

\begin{figure}
\centering
 	\includegraphics[width=0.8\columnwidth]{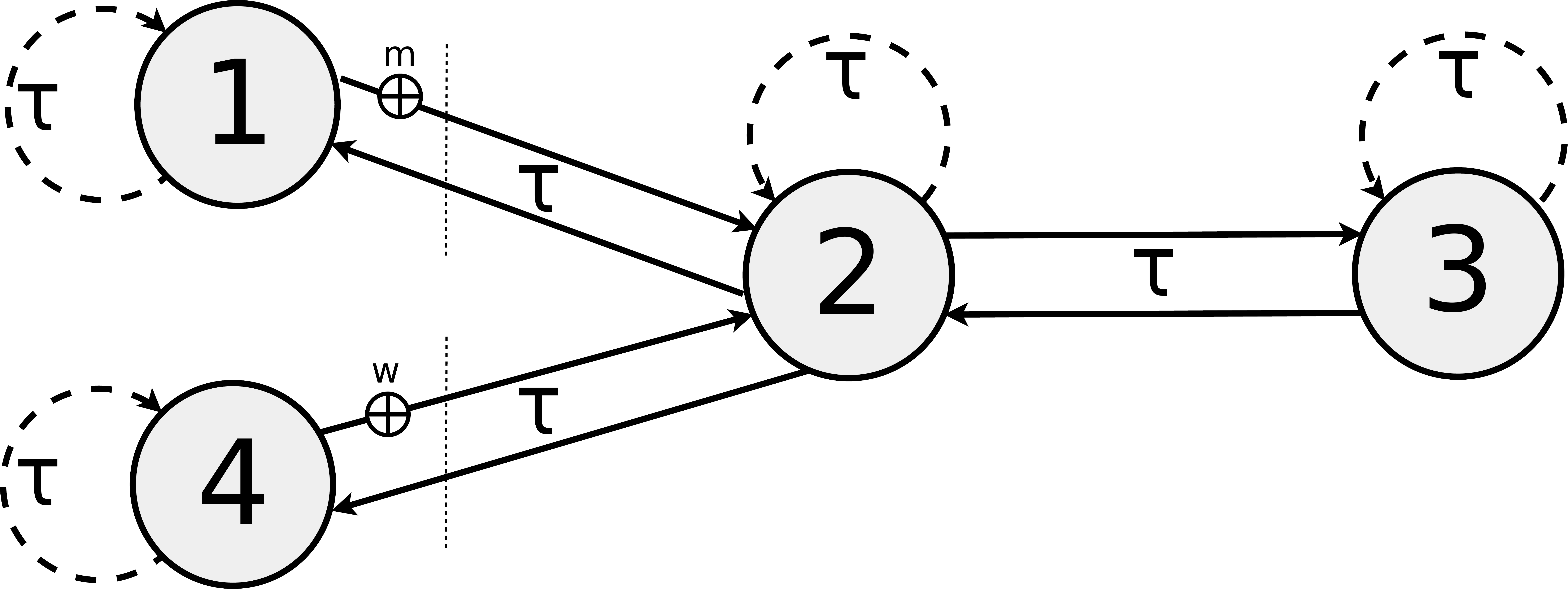}
	\caption{Setup of a star like configuration with unit 2 in the middle. Unit 2 obtains the unperturbed signal of unit 3, and the signal of unit 1 and 4 where a message $m$ and $w$, respectively, is added. Unit 1, 3 and 4 can synchronize perfectly. Hence unit 2 can obtain $m$ and $w$ by subtracting the (unperturbed) signal of unit 3 from the signal of unit 1 and 4, respectively.}
	\label{fig:Setup3Star}
\end{figure}

In terms of chaos communication, unit 2 which obtains the signal of unit 1 and 3 can recover a message, which is added on top of the transmitted signal of unit 1, without any errors.
In the same fashion more units can be added to the system in a star like setup with unit 2 being the relay. The additional units can also add a secret message on top of their transfered signals. Fig.~\ref{fig:Setup3Star} shows the setup for a system with four units, but in principle an arbitrary number of units can be added.
All units of the star, apart from unit 2, synchronize perfectly since they receive an identical driving signal. Unit 2 compares the signal from unit 3 with the incoming signals of units 1 and 4 and perfectly recovers both secret messages. Thus the hub of the star can simultaneously decode any number of secret messages.

\subsection{Four Units Network}

In the following we investigate a ring of four coupled units. For Bernoulli units the synchronization properties for the unperturbed system can be calculated analytically. The stability of synchronization is determined by the eigenvalues of the adjacency matrix which describes the coupling of the units. For an adjacency matrix with the row sum equals one, meaning that all incoming signals are normalized, and in the limit of large delays one finds that a spectral gap between the largest eigenvalue of $\gamma_1=1$ and the second largest eigenvalue $\gamma_2$ is crucial for the stability. Complete synchronization is only possible in the limit of weak chaos if the spectral gap is nonzero \cite{Englert-2011}. For a simple ring network without any self-feedback, where all eigenvalues are $\gamma=1$, no eigenvalue gap exists and it cannot synchronize. Adding an additional link with the coupling strength $\sigma$ changes the eigenvalues such that a gap occurs so that the system is able to synchronize \cite{Kanter-2011}.

The perturbed system, where a noise $m$ is added onto all outgoing signals of unit 1, is depicted in Fig.~\ref{Eqn:4NetworkSys} and is described by following equations
\begin{align}
	x_{t+1}^1 &= (1-\epsilon) f(x_t^1) + \epsilon f(x_{t-\tau}^{4}) \notag \\
	x_{t+1}^2 &= (1-\epsilon) f(x_t^2) + \epsilon f(x_{t-\tau}^{1} +m_{t-\tau}) \notag \\
	x_{t+1}^3 &= (1-\epsilon) f(x_t^3) \notag \\
	& \quad + \epsilon \left( \sigma \, f(x_{t-\tau}^{1} +m_{t-\tau}) + (1-\sigma) \, f(x_{t-\tau}^{2}) \right) \notag \\
	x_{t+1}^4 &= (1-\epsilon) f(x_t^4) + \epsilon f(x_{t-\tau}^{3}) \: .
	\label{Eqn:4NetworkSys}
\end{align}
For such a network the eigenvalue gap in the unperturbed case, and therefor the synchronization ability, is maximal for $\sigma\approx5/8=0.625$. 
Note that for weak chaos and hence for complete synchronization in this network, the chaos of the single units has to be small.
For Bernoulli maps one finds for the critical coupling at which synchronization occurs
\begin{align}
  \epsilon_c \geq \frac{1- 1/a}{1-\abs{\gamma_2}} \:,
\end{align}
with $a$ the parameter of the Bernoulli map. Thus for the maximum eigenvalue gap with $\sigma=0.625$ the system synchronizes for $a\leq1.16$.

\begin{figure}
\centering
 	\includegraphics[width=0.4\columnwidth]{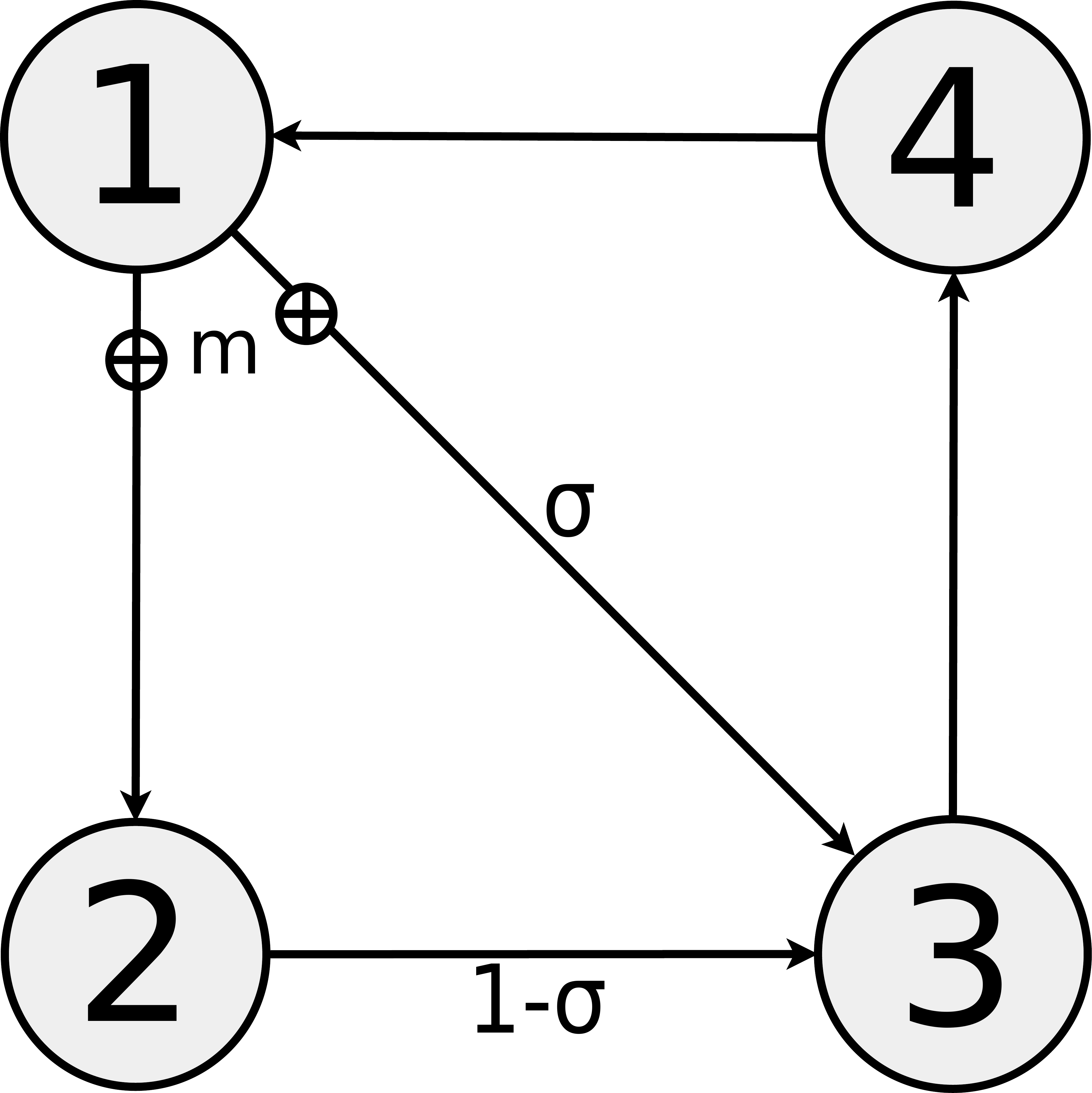}
	\caption{Ring network of four units plus an additional link of strength $\sigma$ in order for the network to synchronize. A perturbation $m$ is added to the outgoing signals of unit 1.}
	\label{fig:Setup4Network}
\end{figure}

\begin{figure}
\centering
 	\includegraphics[width=0.45\textwidth]{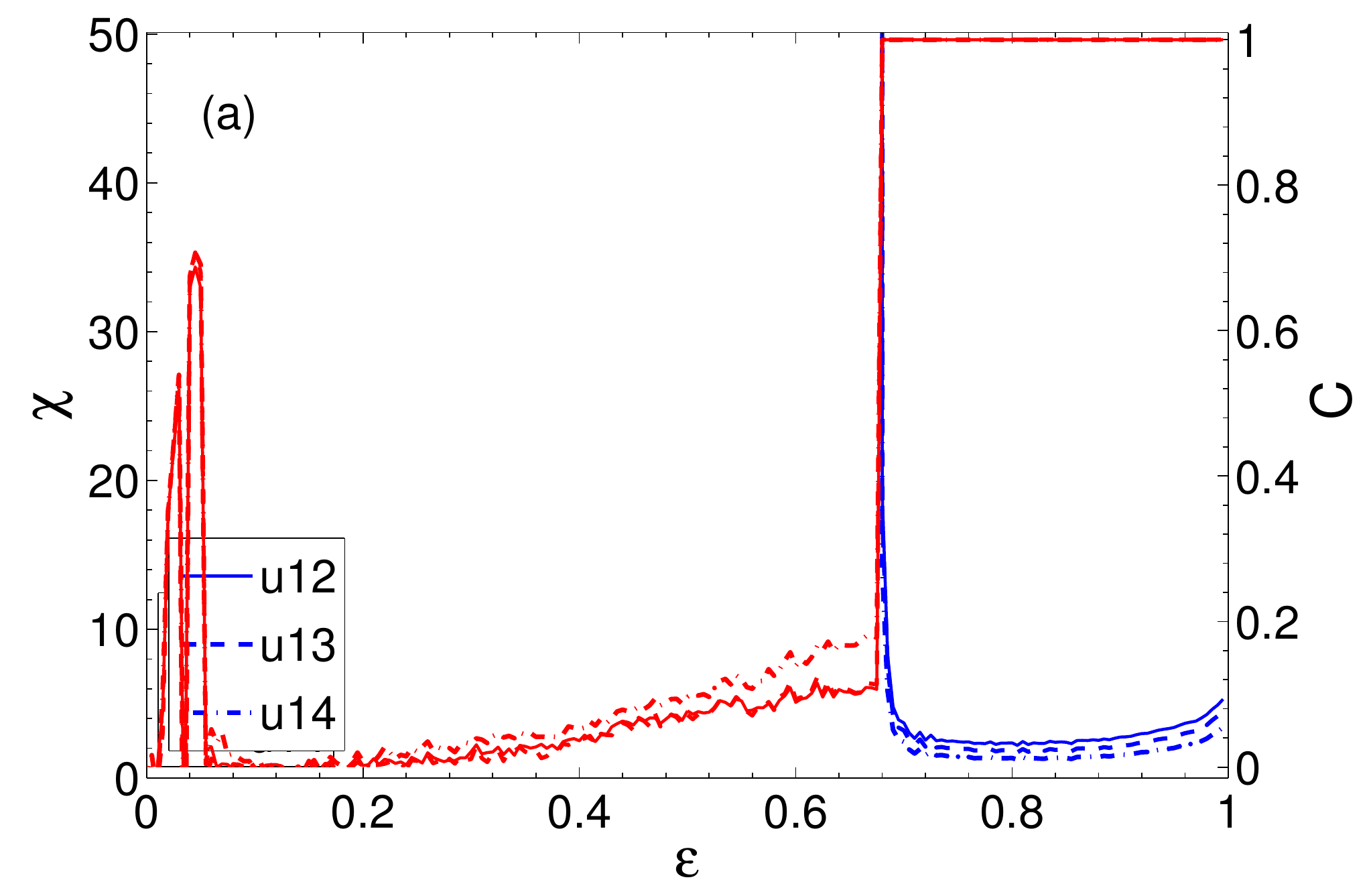}
 	\includegraphics[width=0.45\textwidth]{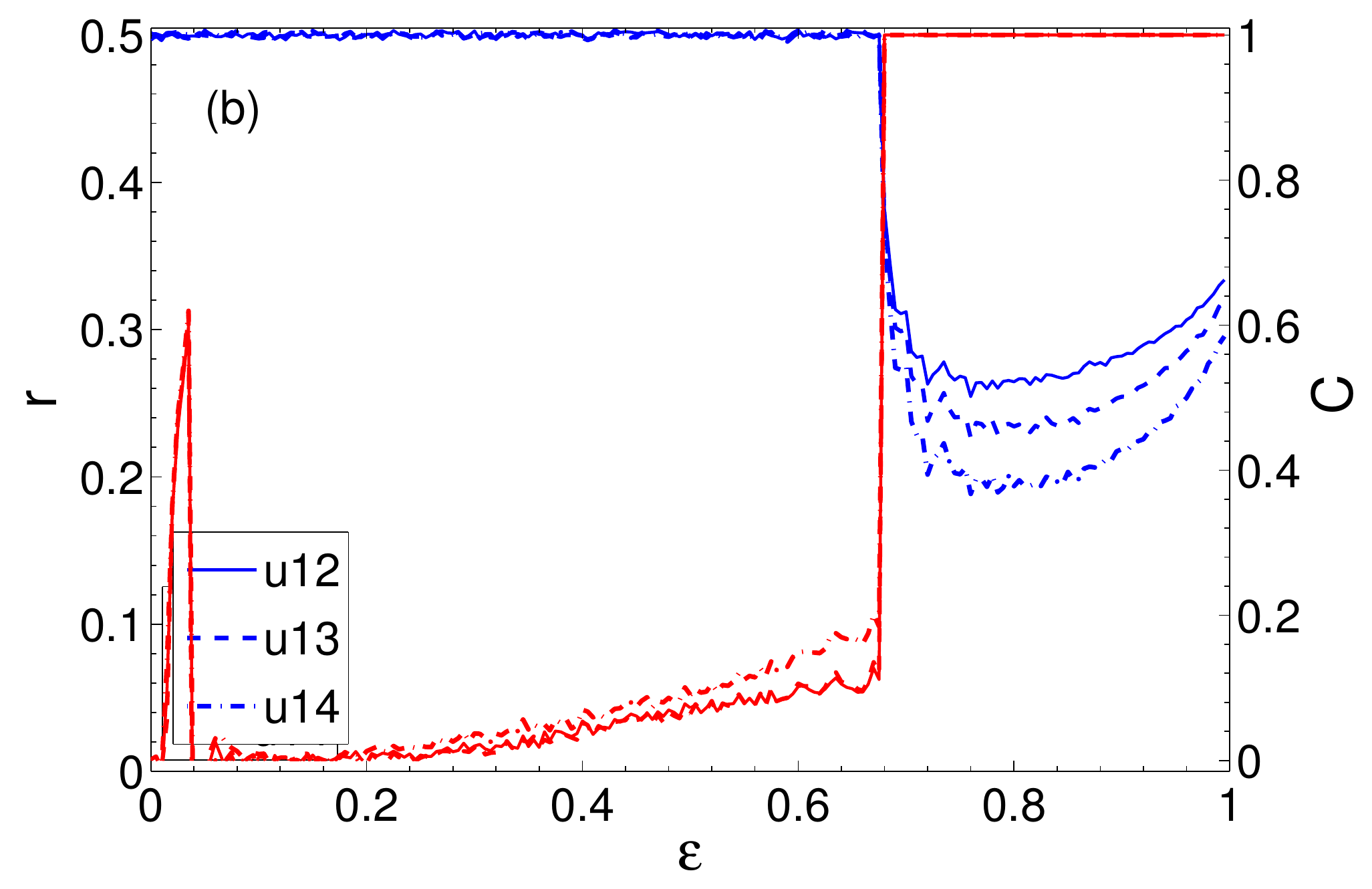}
	\caption{Panel (a) shows second moment $\chi_{1j}$ (blue curves) and cross correlation $C_{1j}$ (red curves) for uniformly distributed random noise, panel (b) shows bit error rate $r_{1j}$ (blue curves) and cross correlation $C_{1j}$ (red curves) for binary random noise as a function of $\epsilon$. Combination of units $ij$ as indicated in the legend. Network of four coupled Bernoulli maps with parameters $a=1.1$, $\tau=100$, $\sigma=0.625$ and $M=10^{-8}$.}
	\label{fig:4NetworkResults}
\end{figure}

For a Bernoulli system the second moment $\chi_{1j}$ and the BER $r_{1j}$ for combination of unit 1 with unit $j=2,3,4$ are shown in Fig.~\ref{fig:4NetworkResults} together with the respective cross-correlations. 
Surprisingly the combination of unit 1 and 2 has the highest second moment and BER whereas unit 4, which is only indirectly driven by unit 1, has the lowest second moment and BER.

\section{Summary}

The linear response of a time-delayed chaotic system to small external perturbations has been studied. The external perturbation drives the system away from the synchronization manifold whereas the dynamics of the system, quantified by transverse Lyapunov exponents, relaxes the system back to the manifold. The competition between these two mechanism results in the linear response studied in this work.

This investigation is motivated by chaos communication, where a secret message is added on top of an exchanged signal between synchronized chaotic units, thus perturbing the system. Nevertheless, the receiver is able to recover the secret message by subtracting its own chaotic trajectory from the received signal. This mechanism has been named \emph{chaos pass filter} since the receiver filters out any external perturbation and essentially responds to the unperturbed chaotic trajectory.

However, our numerical and analytical investigations of iterated maps show that the mechanism of chaos pass filter is much more complex. Perturbations are not just damped, instead the response of the receiver to the perturbation of the sender can be very large. Close to the synchronization transition it diverges and even deep inside the region of synchronization, where the transverse Lyapunov exponents are negative and large, huge excursions away from the synchronization manifold occur resulting in a power law behavior and diverging moments of the distribution of deviations between the sending and receiving unit. Mathematically, this is a consequence of multiplicative and additive noise appearing in the equations of linear response.

The bit error rate of a transmitted binary message is used as another quantity to investigate the linear response. The bit error rate is given by an integral over the distribution of deviations between the trajectories of sender and receiver. For the unsynchronized system the bit error rate is at its maximum of 50\%. Directly at the synchronization transition it decreases to smaller values and shows a complex nonmonotonic behavior inside the region of synchronization which cannot be related to the properties of transverse Lyapunov exponents. 
For special cases we could calculate the bit error rate analytically. Relating it to an iterated function system we found a fractal distribution of deviations yielding a devil's staircase for the bit error rate as a function of model parameters.

The linear response to a periodic perturbation was investigated as well. It shows resonances due to the delayed self-feedback of the sending unit. Depending on parameters and frequency, those resonances can be very large but the response can also be suppressed.

Finally the linear response of a chain of three units and a network of four units has been investigated. We find that for a bi-directionally coupled chain of three units the perturbation is completely filtered out by the unit in the middle and both outer units can synchronize perfectly. Thus the second moment and the bit error rate becomes zero. 

The hub of a star network can receive simultaneously any number of secret messages without any error.

In this work we restricted our investigations to chaotic maps which in some respects have different properties than chaotic flows. But with respect to complete synchronization, maps and flows
are very similar and many of the obtained results are also observed in numerical simulations of chaotic differential equations. Hence we believe that our findings will contribute to a general understanding of linear response of synchronized chaotic systems.

\appendix
\section{Analytic Results For The Bit Error Rate}

Outside the synchronization region the bit error rate trivially equals $1/2$. Inside the synchronization region in general, the bit error rate has to be determined by means of computer simulations. However, for some special cases it can be calculated analytically. 

\subsection{Logistic map, uni-directional setup, $\tau=0$, $\epsilon = 1$, $\kappa = 0$}

In the case of the uni-directional setup with logistic maps and no time delay the bit error rate can be calculated for the point $\epsilon = 1$.
The dynamics is given by 
\begin{equation}
\begin{split}
	x_{t+1} &= f(x_t) \\
	y_{t+1} &= f(x_{t-1} + m_{t-1}) \, ,
\end{split}
\end{equation} 
from which follows that
\begin{equation} \label{eq:dtana}
	d_{t+1} = f'_t m_t \,.
\end{equation} 
From the facts that (1) $f'_t$ and $m_t$ are uncorrelated, (2) the probability distribution of $f'$ is symmetric about $f' = 0$ and (3) $m_t = \pm M$, follows that $d$ has basically the same probability distribution as $f'$. Therefore, the bit error rate can easily be calculated, see Eqs.\ (\ref{eq:ber}) and (\ref{eq:density:fs}):
\begin{equation}
\begin{split}
	r &= \frac{1}{2} \left( 1 - \int\limits_{-M}^{M} p(d) \,\mathrm{d}d \right) = \frac{1}{2} \left( 1 - \int\limits_{-1}^{1} \rho(f') \mathrm{d} f'\right) \\
	&= \frac{\mathrm{arcsec} (4)}{\pi} \approx 0.4196 \; .
\end{split}
\end{equation} 
This is in agreement with the numerical simulations.

\subsection{Logistic map, bi-directional setup, $\tau=0$, $\epsilon = \frac{1}{2}$, $\kappa=0$}

Similarly, in the bi-directional case the bit error rate can be calculated for $\epsilon = \frac{1}{2}$.
The dynamics is given by
\begin{equation}
\begin{split}
	x_{t+1} &= \frac{1}{2} f(x_t) + \frac{1}{2} f(y_t) \\
	y_{t+1} &= \frac{1}{2} f(y_t) + \frac{1}{2} f(x_t + m_t)
\end{split}
\end{equation}
from which follows that
\begin{equation}
	d_{t+1} = \frac{1}{2} f'_t m_t \,.
\end{equation} 
Comparing this with \eq{\ref{eq:dtana}} leads to
\begin{equation}
	r = \frac{1}{2} \left( 1 - \int\limits_{-2}^{2} \rho(f') \mathrm{d} f'\right) = \frac{1}{3}
\end{equation} 
which is also in agreement with the numerical simulations.

\subsection{Bernoulli map, $\tau = 0$}

\begin{figure*}
	\includegraphics[width=0.7\textwidth]{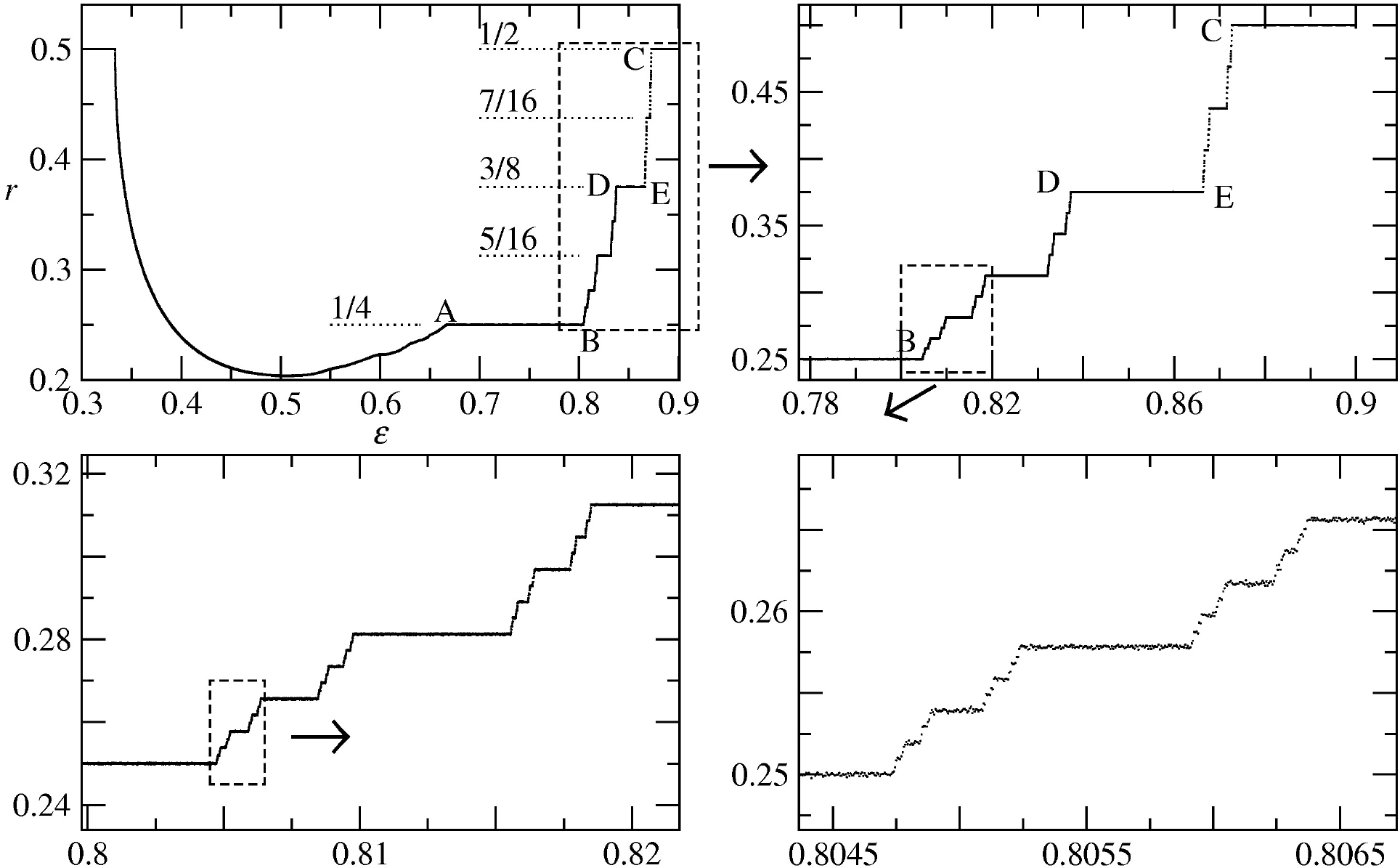}
	\caption{Bit error rate $r$ for uni-directionally coupled Bernoulli maps with $\tau = 0$ and $\kappa = 0$. Zooming in the staircase structure reveals more and more steps.}
	\label{fig:ber:zoom}
\end{figure*}

In \fig{\ref{fig:BER_EpsScan}} (a) one can discover a staircase structure for the bit error rate. For the uni-directional setup this is true for $\epsilon \ge \frac{2}{3}$, while for the bi-directional setup this is valid for $\frac{1}{3} \le \epsilon \le \frac{2}{3}$. For these regions the bit error rate can be calculated analytically. If one takes a closer look at the staircases, \fig{\ref{fig:ber:zoom}}, it becomes apparent that they have infinitely many steps, i.e.\ they are a kind of devil's staircase.

\subsubsection{Uni-directional coupling}

\begin{figure*}
	\subfigure[$\epsilon = 0.6$; \quad $d^-$ and $d^+$ have an overlap in their co-domains.]{
		\includegraphics[width=0.3\textwidth]{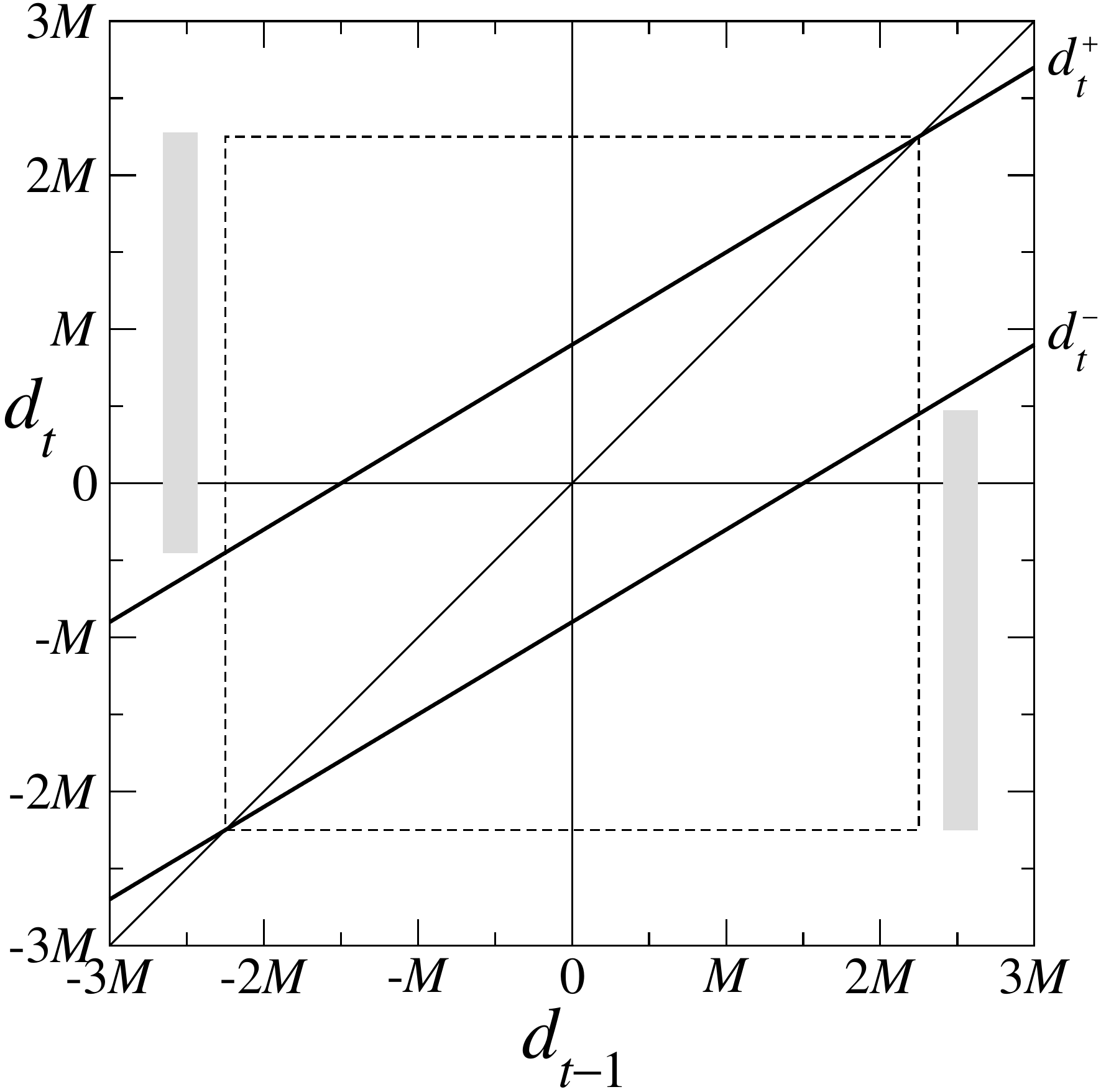}
		\label{fig:linien:a}
	}
	\subfigure[$\epsilon = \frac{2}{3}$]{
		\includegraphics[width=0.3\textwidth]{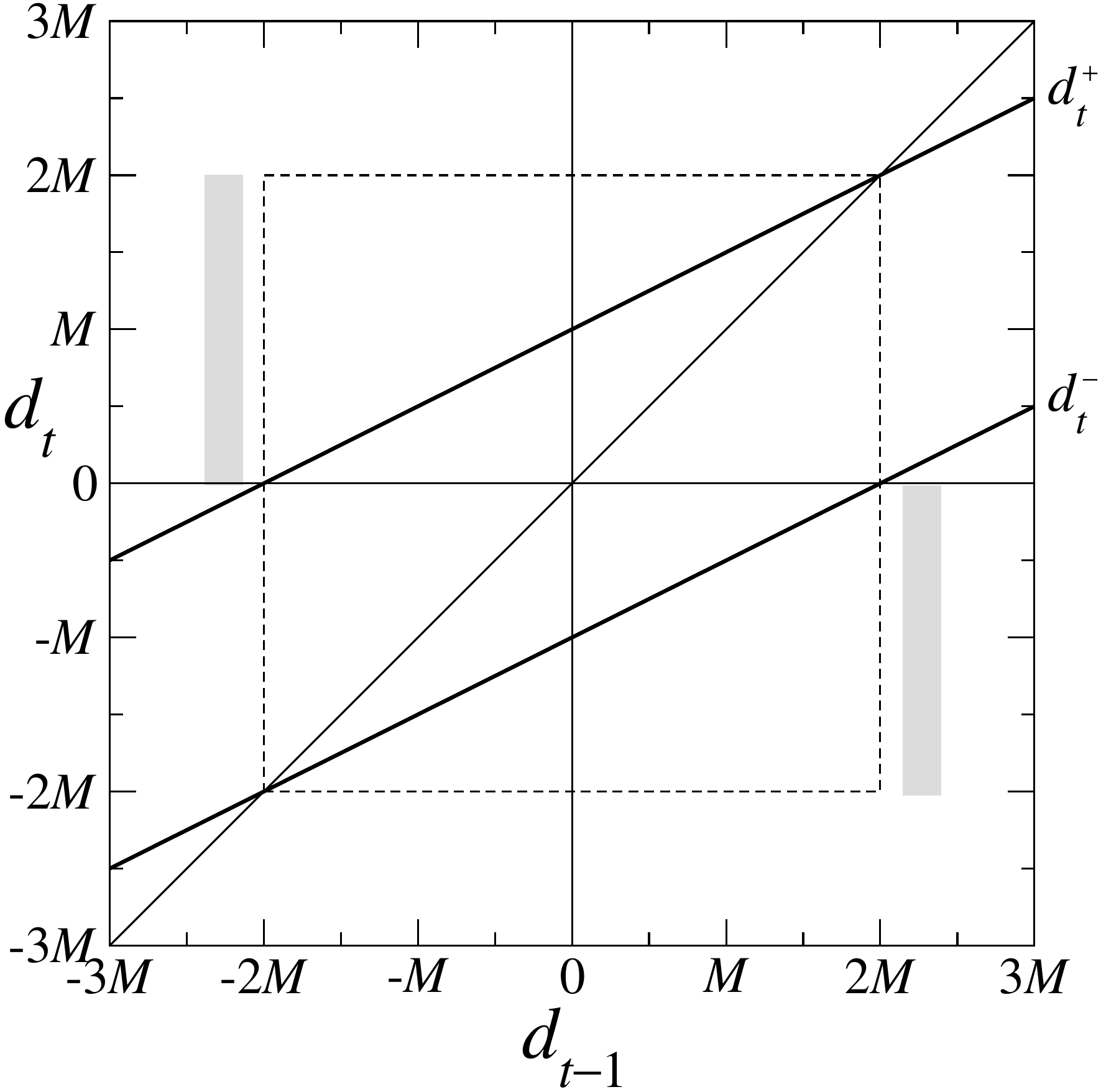}
		\label{fig:linien:b}
	}
	\subfigure[$\epsilon = 0.7$; \quad $d^-$ and $d^+$ have no overlap in their co-domains.]{
		\includegraphics[width=0.3\textwidth]{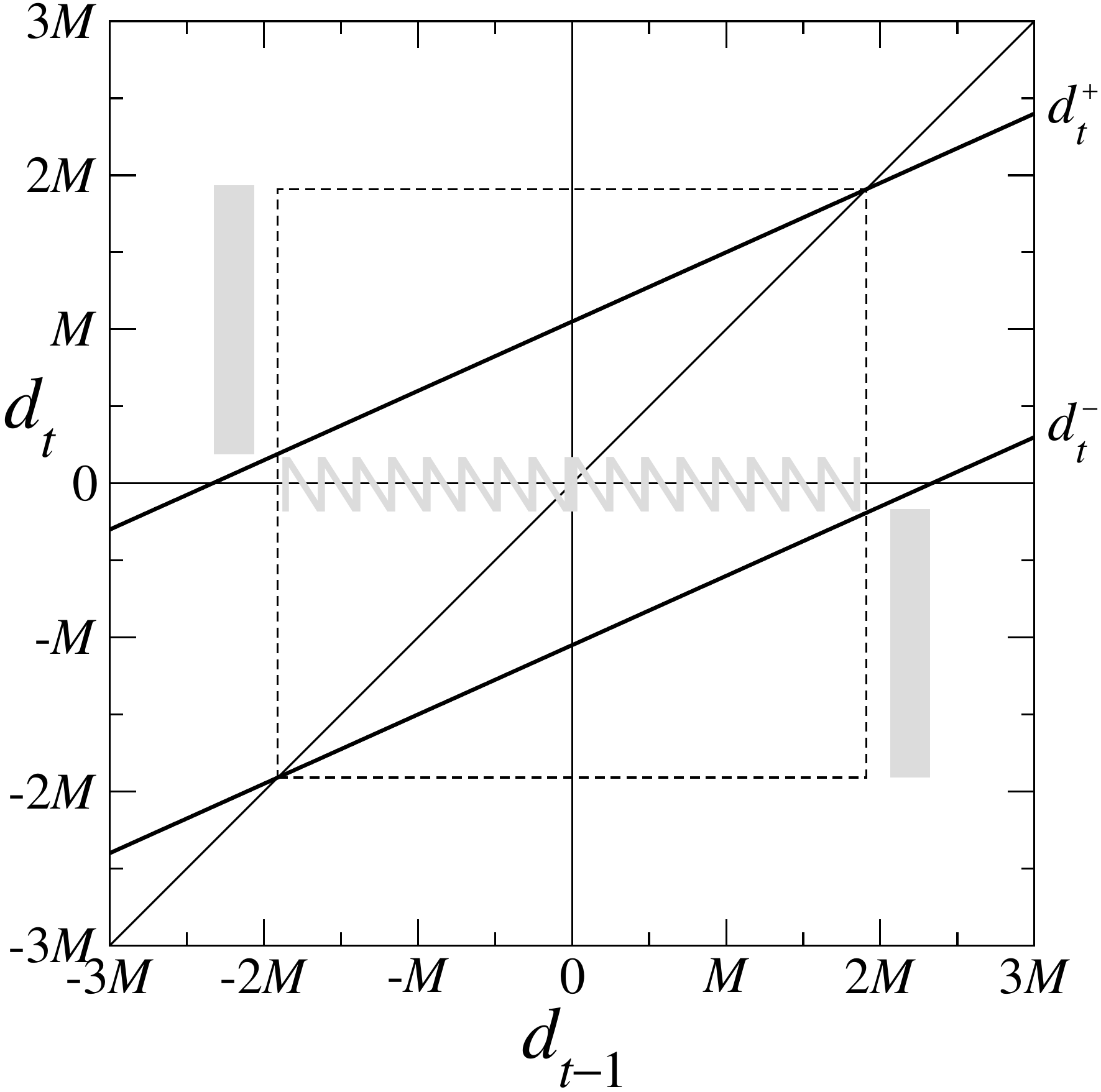}
		\label{fig:linien:c}
	}
	\caption{The iteration $d_t(d_{t-1}) = d_t^{\pm}(d_{t-1})$ of the distances for different coupling parameters $\epsilon$. The dashed boxes represent the interval which $d$ is bounded to due to the fixed points. Additionally, the bisecting line $d_t(d_{t-1}) = d_{t-1}$ is plotted. The gray stripes show the co-domains of $d^-$ and $d^+$.}
	\label{fig:linien}
\end{figure*}

\begin{figure}
	\includegraphics[width=0.4\textwidth]{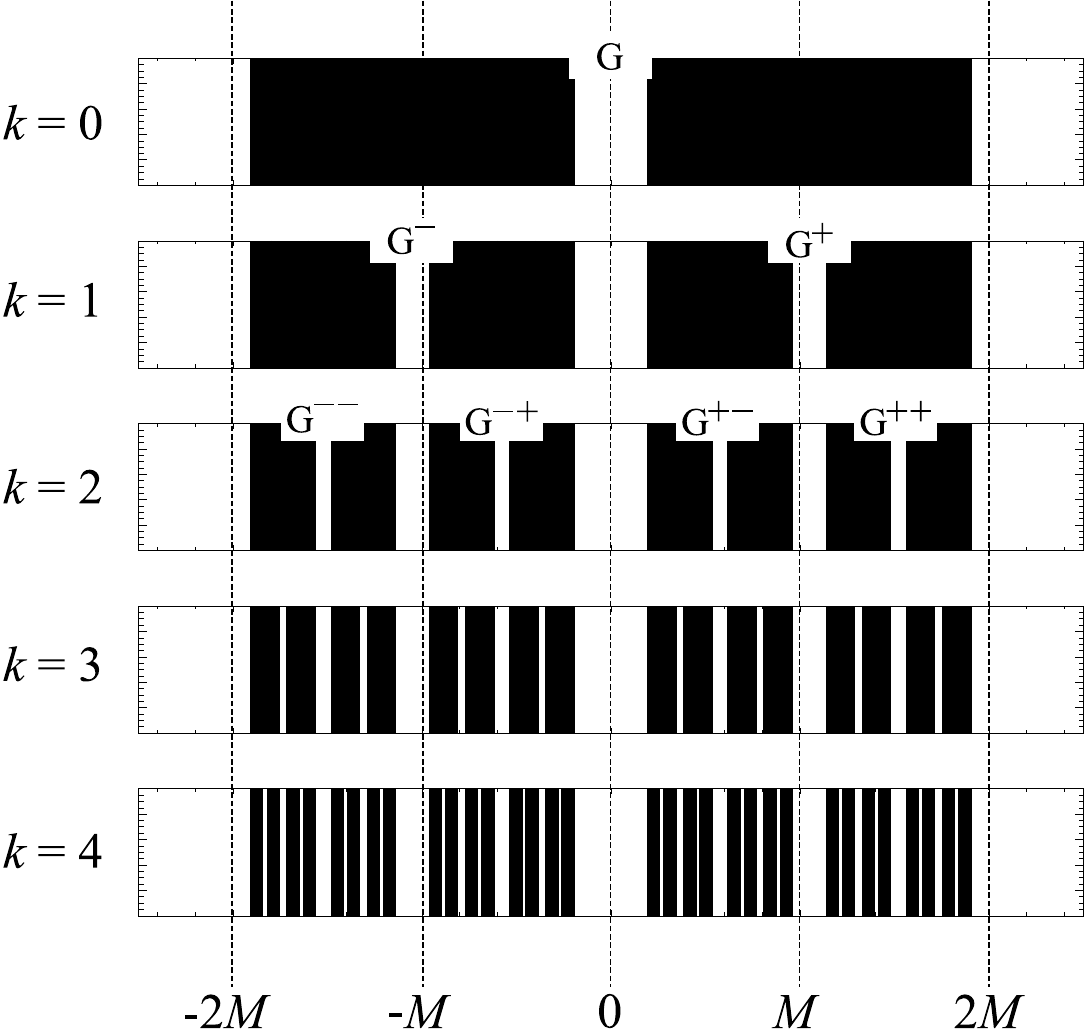}
	\caption{Gaps in the domain/co-domain/distribution of $d^{-}$ and $d^{+}$ for different recursion depths $k$. $\epsilon = 0.7$ .}
	\label{fig:intervalle}
\end{figure}

This staircase structure should be explained here for the case of uni-directional coupling.
The equation for the distance can be written in the following way, see also \eq{\ref{eq:dtEinfach}}:
\begin{equation}
	d_t = 
	\left\{
	\begin{aligned}
		d^{-}_t = \frac{3}{2}(1 - \epsilon) d_{t-1} - \frac{3}{2} \epsilon M \\
		d^{+}_t = \frac{3}{2}(1 - \epsilon) d_{t-1} + \frac{3}{2} \epsilon M
	\end{aligned}
	\right.
\end{equation} 
The two equations represent the two different bits. If $d_t$ is plotted versus $d_{t-1}$, then $d^{-}_t$ and $d^{+}_t$ are two parallel straight lines, see \fig{\ref{fig:linien}}. The values of this iteration $d_t (d_{t-1})$ generate the distribution $p(d)$ from which, in principle, the bit error rate can be calculated. The dashed boxes (-- --) in \fig{\ref{fig:linien}} represent the interval which $d_t$ is bounded to (due to the attracting fixed points).
For $\epsilon < \frac{2}{3}$ the two maps $d^{-}_t$ and $d^{+}_t$ have a certain overlap in their co-domain, see \fig{\ref{fig:linien:a}} (the co-domains are indicated by gray stripes). This fact makes the distribution $p(d)$ complicated. However, for $\epsilon > \frac{2}{3}$ the two maps have no overlap, see \fig{\ref{fig:linien:c}}, and the distribution is manageable analytically. 
As a result of the gap in the co-domain (indicated by a zigzag pattern), a gap in the domain emerges in the next time step. The latter gap produces two further gaps in the co-domain which become gaps in the domain in the next time step. The result of this iterative process is that the distribution $p(d)$ has a fractal support. The first and largest gaps are shown in \fig{\ref{fig:intervalle}}. The first gap is called $G$. The gaps produced by $G$ are $G^-$ and $G^+$. The gaps coming from $G^+$ are called $G^{-+}$ and $G^{++}$; the gaps coming from $G^-$ are called $G^{--}$ and $G^{+-}$ and so on.

Now we want to calculate the exact position of the gaps. The fixed points of $d^-$ and $d^+$ are called $d^{-}_{*}$ and $d^{+}_{*}$. One can easily calculate that
\begin{eqnarray}
	d^-_* &=& -\frac{3\epsilon M}{3\epsilon - 1} \quad\text{ and} \\
	d^+_* &=& +\frac{3\epsilon M}{3\epsilon - 1} \, .
\end{eqnarray} 

From \fig{\ref{fig:linien:c}} it can be seen that 
\begin{equation}
\begin{split} \label{eq:g}
	G &= \left] d^-(d^+_*), d^+(d^-_*) \right[ \\
	  &= \left] -\frac{3\epsilon (3\epsilon - 2) M}{3\epsilon - 1}, \frac{3\epsilon (3\epsilon - 2) M}{3\epsilon - 1} \right[
\end{split}
\end{equation} 
which is about $\left] -0.19 M, 0.19 M \right[$ for $\epsilon = 0.7$, see \fig{\ref{fig:intervalle}}.

The gap $G^+$ is generated by applying $d^+$ to $G$, i.e.
\begin{equation} \label{eq:gplus}
\begin{split}
	G^+ &= \left] d^+(d^-(d^+_*)), d^+(d^+(d^-_*)) \right[ \\
	    &= \left] \frac{3\epsilon (5 - 12\epsilon + 9\epsilon^2) M}{2(3\epsilon - 1)}, -\frac{3\epsilon (5 - 12\epsilon + 9\epsilon^2) M}{2(3\epsilon - 1)} \right[
\end{split}
\end{equation} 
which is about $\left] 0.96 M, 1.14 M \right[$ for $\epsilon = 0.7$, see \fig{\ref{fig:intervalle}}.

Then the gap $G^{-+}$, for example, is generated by applying $d^-$ to $G^+$, i.e.
\begin{equation}
	G^{-+} = \left] d^-(d^+(d^-(d^+_*))), d^-(d^+(d^+(d^-_*))) \right[
\end{equation} 
and so on.

Due to the constant and equal slope of $d^{-}$ and $d^{+}$, and due to the fact that there is no overlap between the co-domains of $d^{-}$ and $d^{+}$, the relative frequency of all distances $d$ which occur (i.e.\ which are not inside a gap) are equal. This means that \fig{\ref{fig:intervalle}} can also be seen as the corresponding histogram; all bars have the same height. 

The bit error rate is related to the integral from $-M$ to $+M$ over the distribution of the distances, see \eq{\ref{eq:ber}}. From \fig{\ref{fig:intervalle}} it can be seen that for $\epsilon = 0.7$ this integral exactly equals $\frac{1}{2}$; thus, the bit error rate equals $\frac{1}{4}$, which is in agreement with \fig{\ref{fig:ber:zoom}}. 

If $\epsilon$ is changed, then the positions of the gaps are changed, too. As long as the gap $G^+$ contains the value $+M$ (= as long as the gap $G^-$ contains the value $-M$), the integral yields $\frac{1}{2}$ and the bit error rate is $\frac{1}{4}$. This explains the plateau AB in \fig{\ref{fig:ber:zoom}}. With the aid of \eq{\ref{eq:gplus}} we can calculate the exact position of this plateau: 
\begin{eqnarray}
&\text{A: } \epsilon =& \frac{2}{3} = 0.\overline{6} \, , \\
&\text{B: } \epsilon =& \frac{1}{3} (1 + \sqrt{2}) \approx 0.804738 \, .
\end{eqnarray} 

Similarly, one gets the point C of \fig{\ref{fig:ber:zoom}}. The bit error rate becomes $\frac{1}{2}$ when the integral starts to be $0$. This is when the gap $G$ is as large as (or larger than) the interval $[-M, M]$. Considering \eq{\ref{eq:g}} yields:
\begin{equation}
\text{C: } \epsilon = \frac{1}{6} (3 + \sqrt{5}) \approx 0.872678 \, .
\end{equation} 

The plateau DE, which has the value $\frac{3}{8}$, can be calculated considering the gap $G^{+-}$. One gets:
\begin{eqnarray}
&\text{D: } \epsilon \approx& 0.837266 \\
&\text{E: } \epsilon \approx& 0.866386
\end{eqnarray} 

All other plateaus can be calculated with the aid of smaller gaps.

\subsubsection{Bi-directional coupling}

The calculations for the case of bi-directional coupling are very similar to the ones for the uni-directional case. Here, only few results should be shown. 

For
\begin{equation}
	0.\overline{3} = \frac{1}{3} \le \epsilon \le \frac{5}{9} = 0.\overline{5}
\end{equation} 
the bit error rate is $0$.

For
\begin{equation}
 0.5749 \approx \frac{1}{6} + \frac{1}{\sqrt{6}} \le 
  \le \frac{2}{3} = 0.\overline{6}
\end{equation} 
the bit error rate is $\frac{1}{4}$.

\clearpage

\end{document}